\documentclass[aps,pre,preprint,groupedaddress,showpacs]{revtex4-2}

\usepackage{graphicx}
\usepackage{color} 

\begin{document}


\title{Contacts, motion and chain-breaking in a two-dimensional granular system displaced by an intruder\\
	\textnormal{Accepted manuscript for Physical Review E, 105, 034903, (2022), DOI: 10.1103/PhysRevE.105.034903, https://doi.org/10.1103/PhysRevE.105.034903}}



\author{Douglas D. Carvalho}
\author{Nicolao C. Lima}
\author{Erick M. Franklin}
 \email{erick.franklin@unicamp.br}
 \thanks{Corresponding author}
\affiliation{%
School of Mechanical Engineering, UNICAMP - University of Campinas,\\
Rua Mendeleyev, 200, Campinas, SP, Brazil\\
}%


\date{\today}

\begin{abstract}
	
We investigate numerically how the motion of an intruder within a two-dimensional granular system affects its structure and produces drag on the intruder. We made use of discrete numerical simulations in which a larger disk (intruder) is driven at constant speed amid smaller disks confined in a rectangular cell. By varying the intruder's velocity and the basal friction, we obtained the resultant force on the intruder and the instantaneous network of contact forces, which we analyze at both the cell and grain scales. We found that there is a bearing network that percolates forces from the intruder toward the walls, being responsible for jammed regions and high values of the drag force, and a dissipative network that percolates small forces within the grains, in agreement with previous experiments on compressed granular systems. In addition, we found the anisotropy levels of the contact network for different force magnitudes and regions, that the force network can reach regions far downstream of the intruder by the end of the intruder's motion, that the extent of the force network decreases with decreasing the basal friction, and that the void region (cavity) that appears downstream the intruder tends to disappear for lower values of the basal friction. Interestingly, our results show that grains within the bearing chains creep while the chains break, revealing the mechanism by which bearing chains collapse.
\end{abstract}


\maketitle


\section{\label{sec:Intro} INTRODUCTION}

The motion of a solid intruder in a granular medium is commonly found in nature and human activities. For example, we find it in the motion of animals and machines over and within granular matter (snakes, worms, vehicles, etc.), the penetration of solid bodies in sand (such as ballistic objects), and the impact of objects on sandy surfaces (such as the landing of space probes on other planets) \cite{Askari, Zheng1}. Depending on velocities, lengths, materials and concentrations involved, the granular system is forced to move in different manners. For instance, the motion can be highly inertial, with great part of the kinetic energy being dissipated through friction and collisions within grains \cite{Bester}, or it can be in quasistatic regime, being dominated by the formation and destruction of contact networks and stick-slip motion \cite{Kolb1, Tordesillas, Kozlowski, Carlevaro}.

When granular materials move in dense and quasistatic regimes, inter-particle forces are transmitted via a history-dependent contact network, leading in many instances to an anisotropic distribution of stresses \cite{Radjai1, Majmudar}. Due to local reorganizations of the granular packing \cite{Houssais}, jamming and unjamming regions appear depending on the persistence or fail of contact networks \cite{Cates, Majmudar, Bi, Seguin2, Behringer_1}, with, respectively, grains resisting to strong external forces or moving under weaker forces. In the case of a granular medium being displaced by an intruder, breaking and formation of force networks result in time varying drag on the intruder \cite{Kolb1, Seguin1}, which gets stronger as packing fractions approach the limit for jamming.

Given their importance, many studies were devoted over the last decades to stress transmission and jamming in granular matter under normal and shear stresses \cite{Radjai1, Cates, Majmudar, Bi, Seguin2}. Radjai et al. \cite{Radjai1} showed that the stress transmission in a two-dimensional (2D) packing of rigid spheres under biaxial compression occurs through two complementary networks: a load-bearing network and a dissipative network. The former is a network of nonsliding contacts that transmit strong forces (higher than the average), carrying the deviatoric load and presenting anisotropy induced by shear, while the latter is a network of sliding contacts that transmit weak forces (smaller than the average), carrying only load contributing to the average pressure and presenting anisotropy in a direction orthogonal to that of the load-bearing network. Later, Seguin \cite{Seguin2} investigated the force network of a monolayer of disks under vibration compressed above the limit for jamming. The results corroborate the existence of load-bearing and dissipative networks, and show that the latter is characterized by cycles consisting of 3 to 8 grains that are linked to the load-bearing chains, relieving part of the load. Cates et al. \cite{Cates} investigated fragile states in colloidal suspensions and granular materials, being defined as those whose internal structure has organized itself to support loads in certain directions, but not in others. They showed that those states result from the formation of force chains aligned in preferential directions, and, therefore, fragile matter undergoes jamming and is able to support loading in such directions, while it undergoes unjamming and suffers plastic deformation in others. Bi et al. \cite{Bi} showed that granular materials sheared by external stresses present not only the isotropic jamming observed in shear-free conditions, but also fragile states and shear jamming that appear at particle fractions lower than those necessary for isotropic jamming. They showed also that the fragile state appears under small shear stresses and is characterized by force chains that are one-directional, while the shear jamming results from stronger shear stresses with a force network that percolates in different directions.

For the case of an intruder moving in granular matter, local variations of particle fraction, forming compressed fronts and expanded trails \cite{Kolb1, Seguin1}, together with the breaking and reorganization of the force networks around the intruder, make the prediction of granular motion and drag forces complex. Many studies were therefore devoted to the drag force on intruders \cite{Albert, Albert2, Stone, Geng, Costantino, Kolb1, Seguin1, Kozlowski, Carlevaro}. In particular, Kolb et al. \cite{Kolb1} investigated experimentally the drag force on the intruder and the motion of grains around it as the intruder was driven within a bidimensional granular system consisting of disks. They showed the formation of a region in front (upstream) of the intruder where grains reach the jamming packing fraction (compression), and a region behind (downstream) the intruder where a cavity without grains (expansion) appears. As the intruder moves, grains recirculate intermittently from the compressed front toward the downstream region with the occurrence of chain breaking and unjamming, making the drag force on the intruder to fluctuate, sometimes very strongly, around a mean value. They showed also that the cavity tends to disappear and the drag to increase greatly as the average particle fraction grows because the compressed front is confined by lateral walls, leaving no room for local compression/expansion in the limit of the highest possible packing fraction. Seguin et al. \cite{Seguin1} inquired further into the motion around an intruder in a granular system similar to that of Ref. \cite{Kolb1}, but using simultaneously photoelastic and tessellation techniques to measure the strain and stress rates at the grain scale. They showed that, although the strain and shear are localized, the macroscopic friction coefficient $\mu$ (ratio of shear to normal stresses) depends on the azimuthal direction, indicating that a local rheology is not adequate to describe the motion of grains around the intruder.

More recently, Kozlowski et al. \cite{Kozlowski} and Carlevaro et al. \cite{Carlevaro} investigated the effects of the packing fraction and the interparticle and basal frictions (the latter between the bottom wall and the grains, excluding the intruder) on the motion of an intruder moving within a bidimensional granular system in a Couette geometry. The experiments \cite{Kozlowski} made use of photoelastic disks moving over either a glass plate or a layer of water, while the numerical simulations \cite{Carlevaro} were 2D and varied the friction coefficient (static and dynamic) over broader ranges. In both, the intruder was driven by the continuous loading of a spring. The experiments showed that in the presence of basal friction there are two regimes depending on the particle fraction: at low particle fractions, an intermittent regime where the intruder moves freely between clogging events appears, while at high particle fractions a stick-slip regime takes place, where the intruder moves through fast slip events alternated with long periods of creep. In the absence of basal friction (water layer), only the intermittent regime is observed. The numerical simulations showed that the intermittent to stick-slip transition is highly affected by the dynamic coefficient of basal friction, with the intermittent regime occurring for values below 0.1 and the stick-slip for higher values, while it is virtually independent of the static coefficient, which contributes mainly to the duration of stick events.

Although extensively studied, the physics involved in a granular medium displaced by an intruder remains to be fully understood and important issues need to be investigated further. This paper presents a numerical investigation of a three-dimensional (3D) cylindrical intruder (disk) driven at constant speed within an assembly of smaller bidisperse disks (3D cylinders) confined in a rectangular cell, with the same configuration of Seguin et al. \cite{Seguin1} (quasistatic regime). We made use of the open-source code LIGGGHTS \cite{Kloss, Berger} and of the DESIgn toolbox \cite{Herman} to perform Discrete Element Method (DEM) computations for an ensemble of disks with static and dynamic coefficients of friction. We first validate our numerical computations by replicating some of the experimental results obtained by Ref. \cite{Seguin1}, and we afterward investigate further the motion of particles and force transmission. We find that there is a bearing network that percolates large forces from the intruder toward the walls, being responsible for jammed regions and high values of the drag force, and a dissipative network that percolates small forces, in agreement with previous observations for compressed 2D granular systems \cite{Radjai1, Seguin2}. In addition, we find the anisotropy levels of the contact network for different force magnitudes and regions, that bearing chains occur preferentially in long chains in front of the intruder (which we associate with local jamming induced by shear), and that the force network can reach regions far downstream of the intruder by the end of the intruder's motion. By varying the coefficients of basal friction, we show that bearing networks transmit stronger forces within longer distances for higher basal friction, and that the void region (cavity) that appears downstream of the intruder tends to disappear for lower values of basal friction. Interestingly, our results show that grains within the bearing chains creep while the chains break, revealing the mechanism by which bearing chains collapse, and allowing the intruder to proceed with its motion.

In the following, Secs. \ref{sec:model} and \ref{sec:setup} present, respectively, the model equations and numerical setup, and Sec. \ref{sec:Res} presents the results for the formation of contact networks, anisotropic levels, creep motion, and drag on the intruder. Section \ref{sec:Conclu} presents the conclusions.

\section{\label{sec:model} MODEL DESCRIPTION}

Our numerical investigation was conducted with {DEM} \cite{Cundall}, where the dynamics of each individual particle was computed using the forces and torques on each of them. We used the open-source code LIGGGHTS \cite{Kloss, Berger} for DEM computations, and, in order to produce disks that have friction with the bottom and lateral walls and between them, we made use of the DESIgn toolbox developed by Herman \cite{Herman}.

The dynamics of each particle is computed by the linear and angular momentum equations, given by Eqs. \ref{Fp} and \ref{Tp}, respectively,

\begin{equation}
	m\frac{d\vec{u}}{dt}= \vec{F}_{c} + m\vec{g}
	\label{Fp}
\end{equation}

\begin{equation}
	I\frac{d\vec{\omega}}{dt}=\vec{T}_{c}
	\label{Tp}
\end{equation}

\noindent where $\vec{g}$ is the acceleration of gravity and, for each particle, $m$ is the mass, $\vec{u}$ is the velocity, $I$ is the moment of inertia, $\vec{\omega}$ is the angular velocity, $\vec{F}_{c}$ is the resultant of contact forces between solids, and $\vec{T}_{c}$ is the resultant of contact torques between solids. The contact forces and torques are computed by Eqs. \ref{Fc} and \ref{Tc}, respectively,

\begin{equation}
	\vec{F}_{c} = \sum_{i\neq j}^{N_c} \left(\vec{F}_{c,ij} \right) + \sum_{i}^{N_w} \left( \vec{F}_{c,iw} \right)
	\label{Fc}
\end{equation}

\begin{equation}
	\vec{T}_{c} = \sum_{i\neq j}^{N_c} \vec{T}_{c,ij} + \sum_{i}^{N_w} \vec{T}_{c,iw}
	\label{Tc}
\end{equation}

\noindent where $\vec{F}_{c,ij}$ and $\vec{F}_{c,iw}$ are the contact forces between particles $i$ and $j$ and between particle $i$ and the wall, respectively, $\vec{T}_{c,ij}$ is the torque due to the tangential component of the contact force between particles $i$ and $j$, and $\vec{T}_{c,iw}$ is the torque due to the tangential component of the contact force between particle $i$ and the vertical wall. $N_c$ - 1 is the number of particles in contact with particle $i$, and $N_w$ the number of particles in contact with the wall. Since the grains are disks lying on a horizontal wall, $\vec{F}_{c,iw}$ includes the friction force between the bottom wall and each grain, which follows the Coulomb law with static and dynamic values.

For the contact forces between particles ($\vec{F}_{c,ij}$), and between particles and the lateral walls (included in $\vec{F}_{c,iw}$), the elastic Hertz-Mindlin contact model \cite{direnzo} is used. This model consists in the combination of two spring-dashpots, the first one including the normal interactions and a Coulomb friction coefficient, and the second one including the tangential forces. Equations for computing the normal and tangential forces based on particle overlaps (modeling deformations) are available in the Supplemental Material \cite{Supplemental}.

Because the DESIgn toolbox originally does not compute the friction between the grains and the bottom wall (included in $\vec{F}_{c,iw}$), we implemented that in its library. The friction force was modeled in a similar manner as in Ref. \cite{Carlevaro}: if a grain $i$ is moving at a speed $v_{i} = |\vec{u}_{i}|$ above a threshold value $v^{\prime}$ ($v_{i} > v^{\prime}$), then a dynamic friction force with the bottom wall is considered as being $\vec{F}_{c,iw}$ = $-\mu_{d,g}m_{i}|\vec{g}|\vec{u}_{i}/|\vec{u}_{i}|$. Conversely, if it is moving with a velocity $v_{i}$ smaller than or equal to the threshold value $v^{\prime}$ ($v_{i} \leq v^{\prime}$), then a static friction force with the bottom wall $\vec{F}_{c,iw}$ = $-\mu_{s,g}m_{i}|\vec{g}|\vec{u}_{i}/|\vec{u}_{i}|$ is applied and the particle is stopped by setting $v_{i}$ = 0. This ensures that a grain will only resume its motion if the forces exerted by the other grains exceed the static friction force \cite{Carlevaro}. In this model, we do not consider rotational friction between the grains and the bottom wall.

\section{\label{sec:setup} NUMERICAL SETUP}

\begin{figure}[h!]
	\begin{center}
	\begin{minipage}{0.49\linewidth}
		\begin{tabular}{c}
			\includegraphics[width=0.80\linewidth]{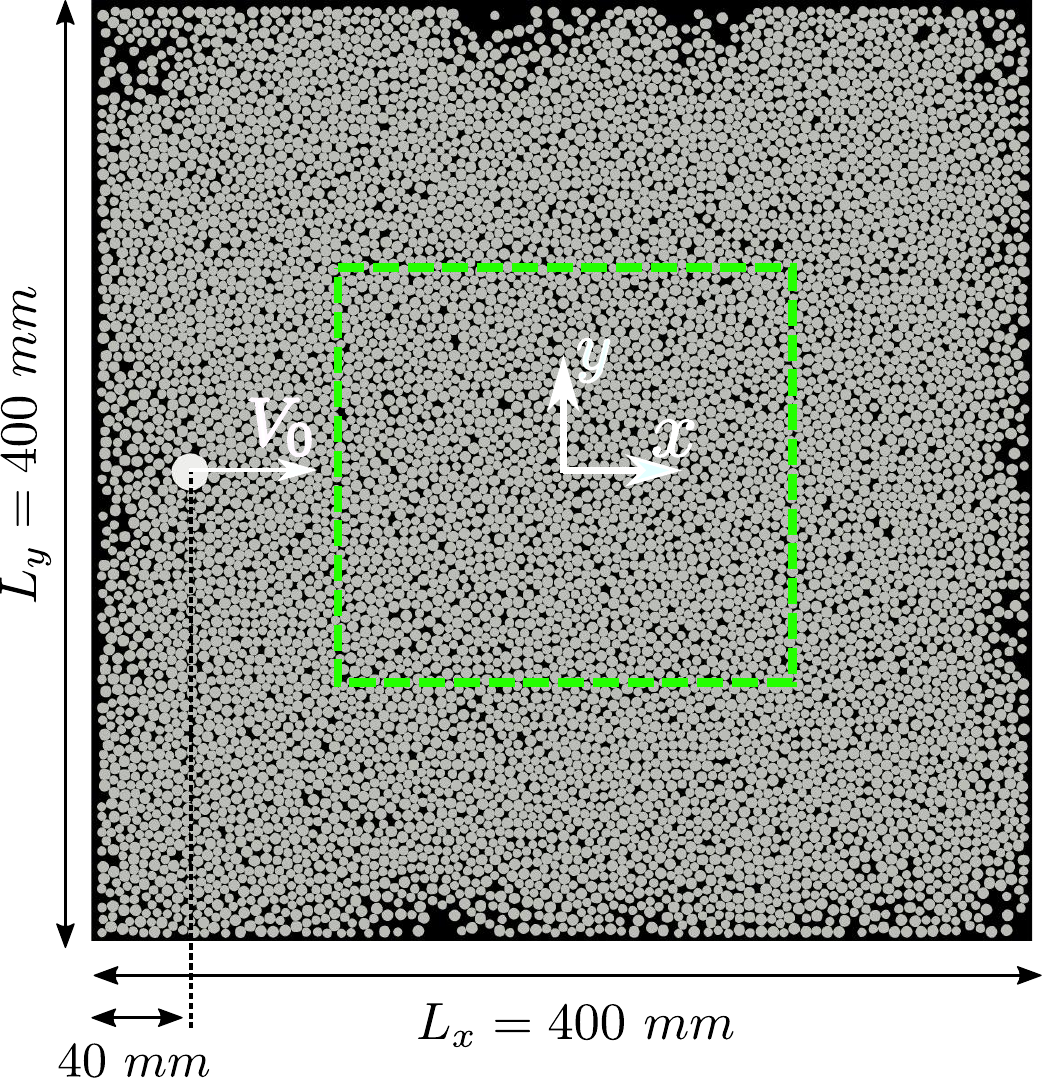}\\
			(a)
		\end{tabular}
	\end{minipage}
	\hfill
	\begin{minipage}{0.49\linewidth}
		\begin{tabular}{c}
			\includegraphics[width=0.7\linewidth]{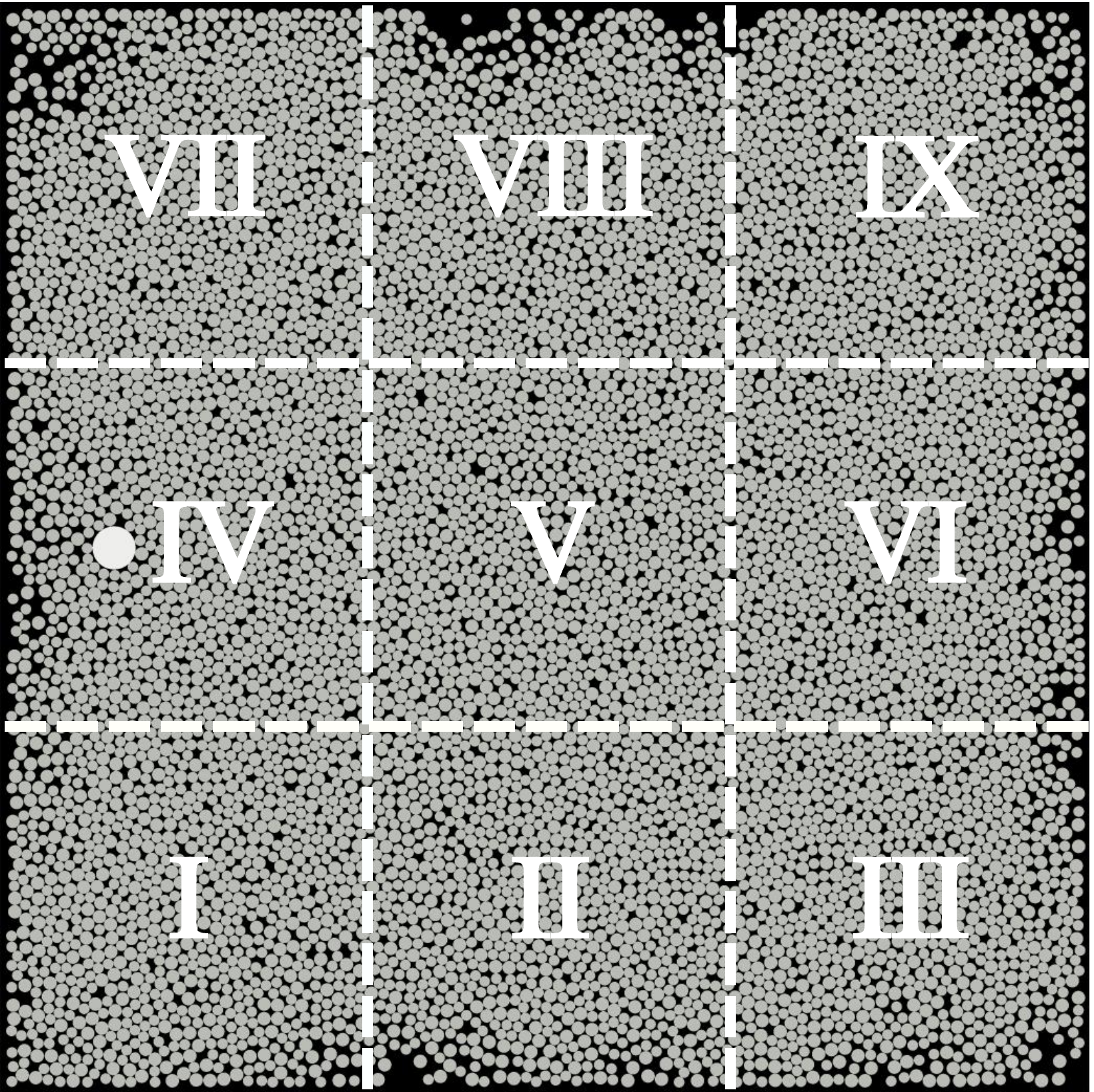}\\
			(b)
		\end{tabular}
	\end{minipage}
	\hfill
	\begin{minipage}{0.49\linewidth}
		\begin{tabular}{c}
			\\
			\includegraphics[width=0.80\linewidth]{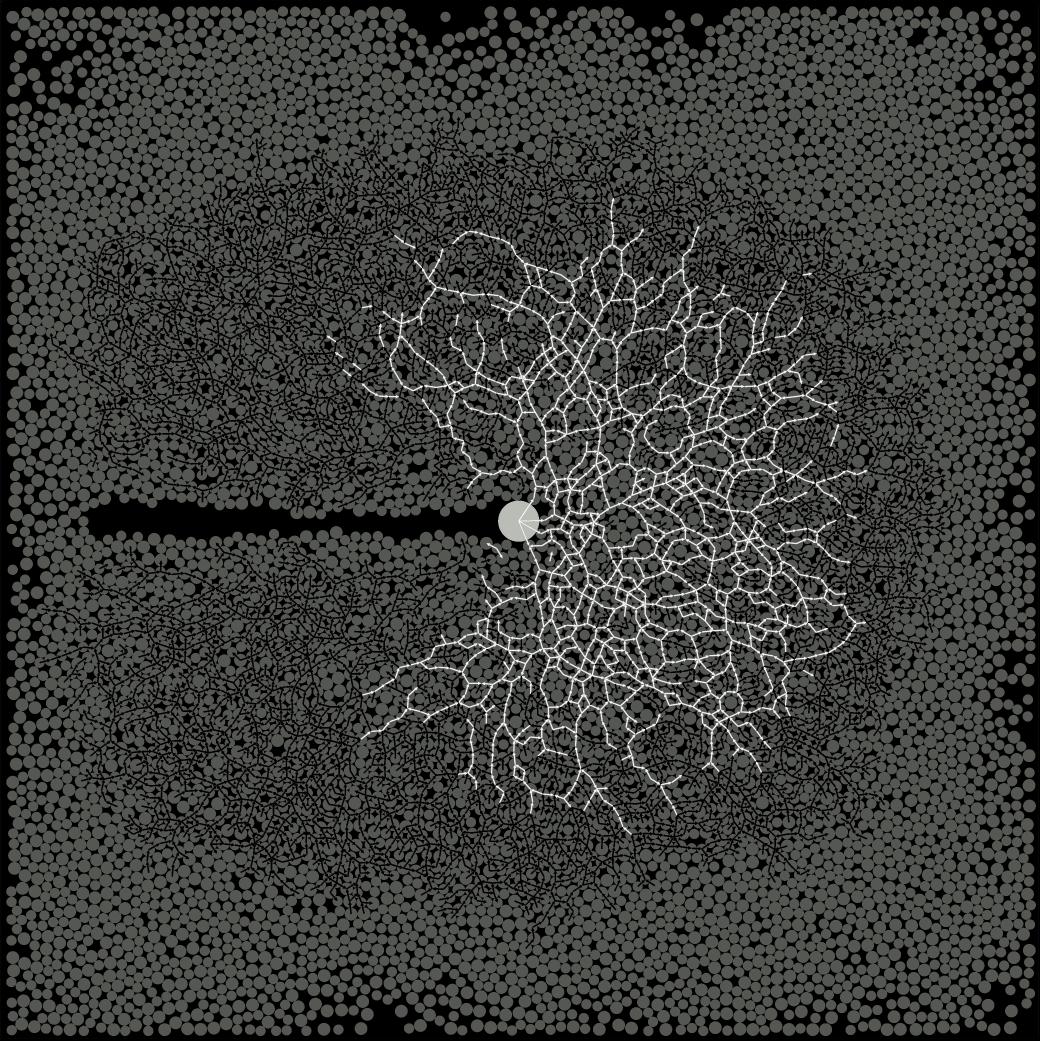}\\
			(c)
		\end{tabular}
	\end{minipage}
	\hfill
	\begin{minipage}{0.49\linewidth}
		\begin{tabular}{c}
			\\
			\includegraphics[width=0.8\linewidth]{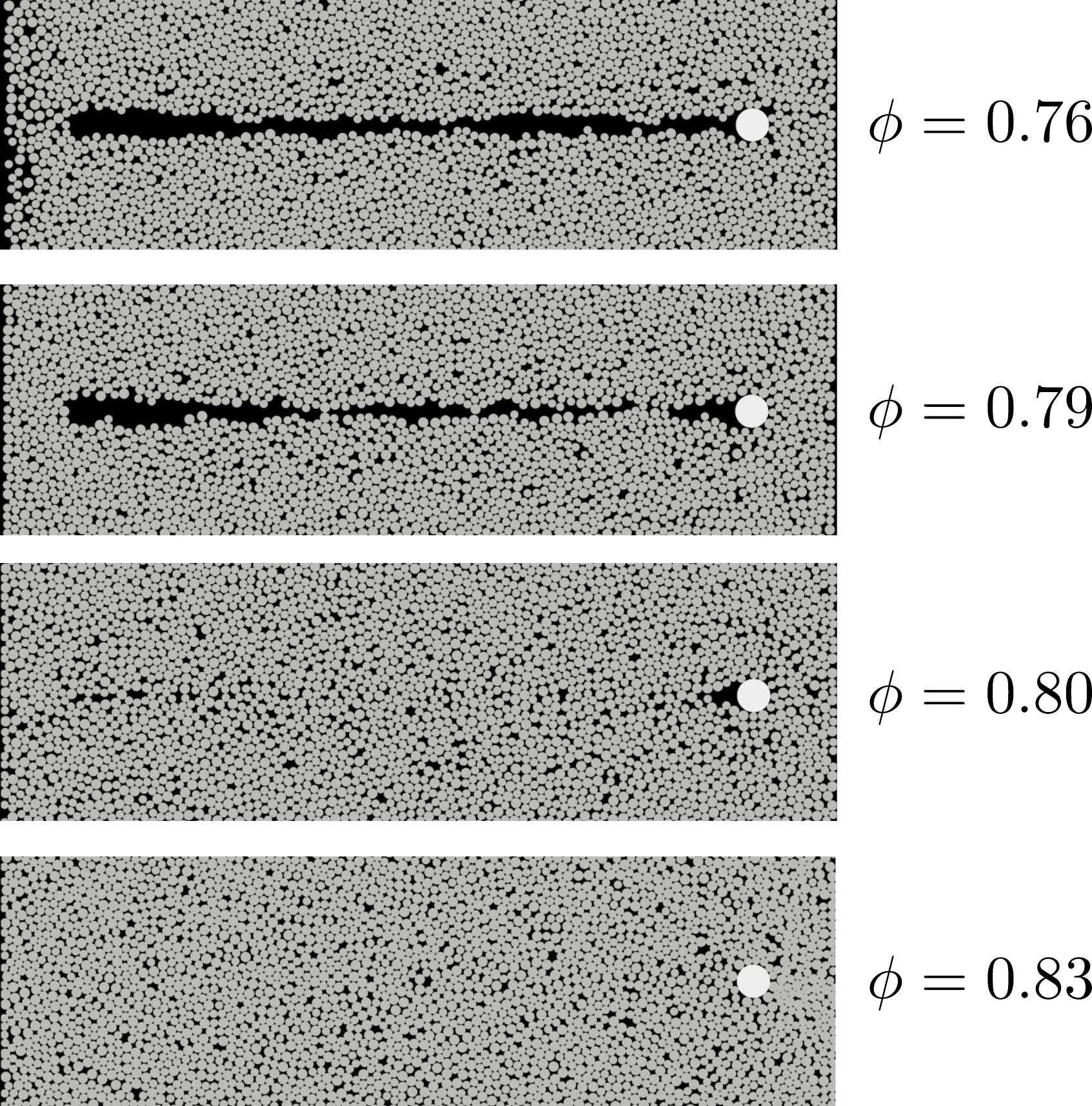}\\
			(d)
		\end{tabular}
	\end{minipage}
	\hfill
\end{center}
	\caption{(a) Numerical setup for $\phi$ = 0.76. (b) Different sub-regions of the entire domain analyzed individually. (c) Force chains formed during the motion of the intruder. Clear networks correspond to bearing (stronger) chains and darker networks to dissipative (weaker) chains. (d) Cavity formed downstream of the intruder for different packing fractions.}
	\label{fig:setup}
\end{figure}

The computed system consisted basically of an assembly of 3D disks settled over a horizontal wall and confined by vertical walls, and of a larger 3D solid disk (intruder) that moved at constant velocity amid the other disks (the top wall was absent). Although the solid objects are 3D disks, we employ the terminology \textit{two-dimensional granular system} since they form a monolayer of particles. The dimensions and properties are roughly the same as in Ref. \cite{Seguin1}, the steel intruder having diameter and height of $d_{int} = 16$ mm and $h_{int} = 3.6$ mm, respectively, and the granular system consisting of a bidisperse mixture of polyurethane (PSM-4) disks with small and large diameters of $d_{s} = 4$ mm and $d_{l} = 5$ mm, respectively (in order to prevent crystallization \cite{Speedy}), and height $h_{g} = 3.2$ mm. We forced the intruder to move within the disks at a constant velocity that varied within 10$^{-1}$ mm/s $\leq$ $V_{0}$ $\leq$ 10 mm/s. The disks were distributed in a proportion of $N_{l}/N_{s} \approx 0.64$, where $N_{s}$ and $N_{l}$ are the numbers of small and large particles, respectively, in a way that the areas occupied by the small and large grains were almost the same. The disks were placed over a horizontal glass plate and were enclosed by vertical glass walls, so that the system dimensions were of $L_x$ $\times$ $L_{y}$ = 400 mm $\times$ 400  mm, where $L_x$ and $L_y$ are the longitudinal and transverse lengths, respectively. All simulations were performed with a fixed cell size (total domain), in a way that the mean packing fraction is kept constant for each computed case, being varied within 0.76 $\leq$ $\phi$ $\leq$ 0.83 by varying the number of disks in each tested case. The number of disks and the corresponding packing fractions are available in the Supplemental Material \cite{Supplemental}, and an example of setup of one simulation can be seen in Fig. \ref{fig:setup}a.

Although the Young's modulus of the steel is $E = 1.96 \times 10^{11}$ Pa, we used a value two orders of magnitude smaller in the numerical simulations in order to decrease the time step without considerably affecting the accuracy of the results \cite{Lommen}. The properties of the materials used in the numerical simulations are summarized in Tab. \ref{tabmaterials}. We do not consider any motion in the direction perpendicular to the $xy$ plane, so that there is no collision between the disks and bottom wall. The intruder is placed initially at the location $x_{i}$ = -160 mm , $y_{i}$ = 0 mm, in the left side of the simulation cell, and is moved at a constant velocity $V_{0}$ from left to right, through the granular medium, toward its final position at $x_{i}$ = 160 mm , $y_{i}$ = 0 mm (Fig. \ref{fig:setup}a). Therefore, for all values of $V_{0}$ and $\phi$ used in the simulations, the intruder traveled a total distance equal to $\Delta S = 320$ mm. The drag force $F_D$ exerted by the grains onto the intruder, the forces on each grain, their displacements, and the contact network are computed at every time step. We defined a region of interest $(ROI)$ of size $A_{ROI}$ = 160 mm $\times$ 160 mm in the center of the domain (green-dashed area in Fig. \ref{fig:setup}a) for computing time averages while avoiding intense boundary effect (see the Supplemental Material \cite{Supplemental} for a figure showing the effect of considering the entire domain on time averages). The remaining computations considered the entire domain. 

\begin{table}[!h]
	\centering
	\caption{Properties of materials used in the simulations: $E$ is the Young's modulus, $\nu$ is the Poisson ratio, $\rho$ is the material density, and $d$ is the particle diameter.}
	\label{tabmaterials}
	\begin{tabular}{l|c|c|c|c|c}
		\hline
		& \textbf{Material} & \textbf{$E$ (Pa)} & \textbf{$\nu$} & \textbf{$\rho$ (kg/m$^{3}$)}& \textbf{$d$ (mm)}\\
		\hline
		Intruder & Steel\footnotesize{$^{(1)}$} & $1.96 \times 10^{9}$  & 0.29 & 7800 & $d_{int} =$ 16            \\
		Grains & Polyurethane\footnotesize{$^{(1),(2)}$} & $4.14 \times 10^{6}$   & 0.50 & 1280 & $d_{s}$ = 4; $d_{l}$ = 5           \\
		Walls & Glass\footnotesize{$^{(1)}$} & $0.64 \times 10^{11}$ & 0.23 & 2500 & $L_x$ = 400; $L_{y}$ = 400\\    
		\hline
		\multicolumn{3}{l}{\footnotesize{$^{(1)}$ Hashemnia and Spelt \cite{Hashemnia}}}\\
		\multicolumn{3}{l}{\footnotesize{$^{(2)}$ Gloss \cite{Gloss}}}
	\end{tabular}
\end{table}

Values for the coefficient of restitution were 0.3 for the grain-grain and  0.7 for the grain-wall and grain-intruder interactions \cite{Gondret, Hashemnia}. For the friction coefficients, we considered only the dynamic coefficient for interactions occurring on surfaces oriented in the vertical plane (grain-grain, intruder-grain and grain-vertical wall interactions), for which we applied the Hertz-Mindlin contact model. Values for the polyurethane-polyurethane and polyurethane-steel found in the literature are relatively high \cite{Hashemnia, Carlevaro} when compared to the other materials involved. The friction between both the intruder and disks with the bottom wall was implemented by ourselves, for which we considered both the static and dynamic coefficients. For that, we defined a threshold velocity $v'$ = 5 $\times$ 10$^{-4}$ m/s for the transition between static and dynamic conditions. Values of the coefficients of restitution (ratio between the momentum just after and prior collision) and friction (Coulomb law) and the threshold velocity used in the simulations are listed in Tab. \ref{tabcoefficients}. Most of the coefficients were obtained from the literature \cite{Carlevaro, Hashemnia, Gondret}, and sensitivity tests varying the coefficients are available in the Supplemental Material \cite{Supplemental}. 

\begin{table}[!h]
	\centering
	\caption{Coefficients and threshold used in the numerical simulations.}
	\label{tabcoefficients}
	\begin{tabular}{l|c|c}
		\hline
		\textbf{Coefficient}  & \textbf{Symbol} & \textbf{Value} \\
		\hline		  
		Restitution coefficient (grain-grain) & $\epsilon_{gg}$ & 0.3 \\
		Restitution coefficient (grain-intruder)\footnotesize{$^{(2)}$} & $\epsilon_{gi}$ & 0.7 \\
		Restitution coefficient (grain-wall)\footnotesize{$^{(3)}$} & $\epsilon_{gw}$ & 0.7 \\
		Dynamic friction coefficient (grain-grain)\footnotesize{$^{(1)}$} & $\mu_{gg}$ & 1.2 \\
		Dynamic friction coefficient (grain-intruder)\footnotesize{$^{(2)}$} & $\mu_{gi}$ & 1.8 \\
		Dynamic friction coefficient (intruder-bottom wall) & $\mu_{iw}$ & 0.7 \\  
		Dynamic friction coefficient (grain-walls)\footnotesize{$^{(1)}$} & $\mu_{gw}$ & 0.4 \\		
		Static friction coefficient (grain-bottom wall) & $\mu_{s,gw}$ & 0.7 \\
		Threshold velocity (dynamic/static friction) & $v^{\prime}$ & $v'$ = 5 $\times$ 10$^{-4}$ m/s\\
		\hline
		\multicolumn{3}{l}{\footnotesize{$^{(1)}$ Carvelaro et al. \cite{Carlevaro}}} \\
		\multicolumn{3}{l}{\footnotesize{$^{(2)}$ Hashemnia et al. \cite{Hashemnia}}} \\
		\multicolumn{3}{l}{\footnotesize{$^{(3)}$ Gondret et al. \cite{Gondret}}}
	\end{tabular}
\end{table}

With the total domain and the particle fraction to be simulated defined, the set of disks with the desired proportion is generated. First, the particles are randomly distributed over a square space larger than the computational domain. Afterward, the space occupied initially by the disks is compressed from the external limits toward its interior until reaching the size of the computational domain. In this initialization process, the number of generated particles is the necessary to achieve the desired packing fraction $\phi$, according to Eq. \ref{eqnphi} \cite{Kolb1}.

\begin{equation}
	\phi = \frac{\frac{\pi}{4}(N_{s}d_{s}^{2} + N_{l}d_{l}^{2})}{L_{x}L_{y} - \frac{\pi}{4}d_{int}^{2}}
	\label{eqnphi}
\end{equation}

\noindent This initialization is necessary because the software is not capable of randomly inserting disks at high particle fractions in the domain within reasonable times (it takes much greater times than those of simulations themselves). Finally, the disks are allowed to relax and, afterward, the simulation starts by setting the intruder into motion at a constant speed. All computations were performed with a time step $\Delta t = 3.2 \times 10^{-6}$ s, which, in the worst scenario, is less than 10 \% of the Rayleigh time (timescale for Rayleigh waves resulting from collisions, given by $t_R$ $=$ $\pi \left( d/2 \right) \left(\rho /G \right)^{1/2} \left( 0.163 \nu + 0.8766 \right)^{-1}$, where $G$ is the shear modulus) \cite{Derakhshani}.

\section{\label{sec:Res} RESULTS AND DISCUSSION}

\subsection{\label{sec:drag} Drag force on the intruder}

\begin{figure}[h!]
	\begin{center}
			\begin{minipage}{0.49\linewidth}
			\begin{tabular}{c}
				\includegraphics[width=0.80\linewidth]{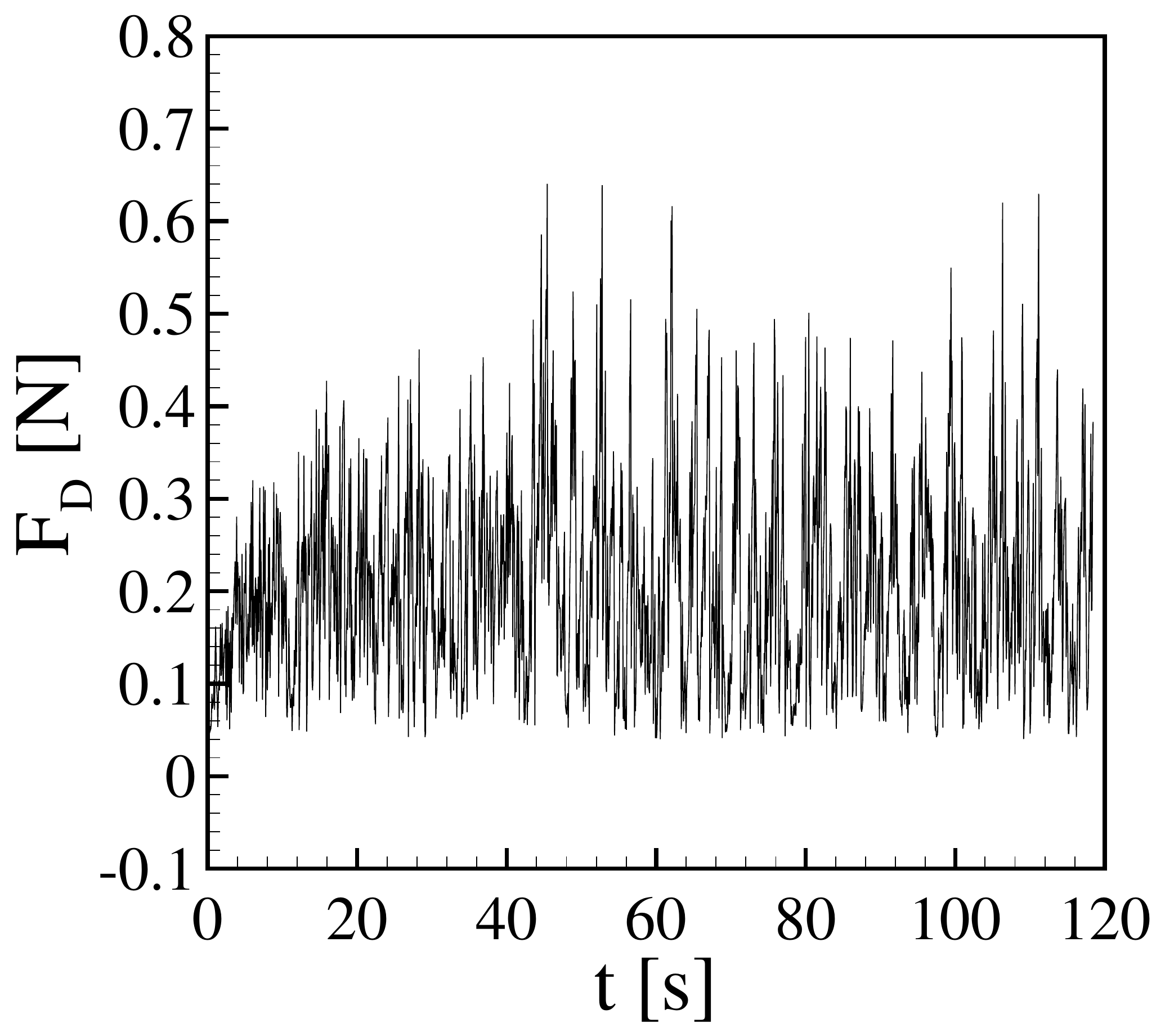}\\
				(a)
			\end{tabular}
		\end{minipage}
		\hfill
		\begin{minipage}{0.49\linewidth}
			\begin{tabular}{c}
				\includegraphics[width=0.80\linewidth]{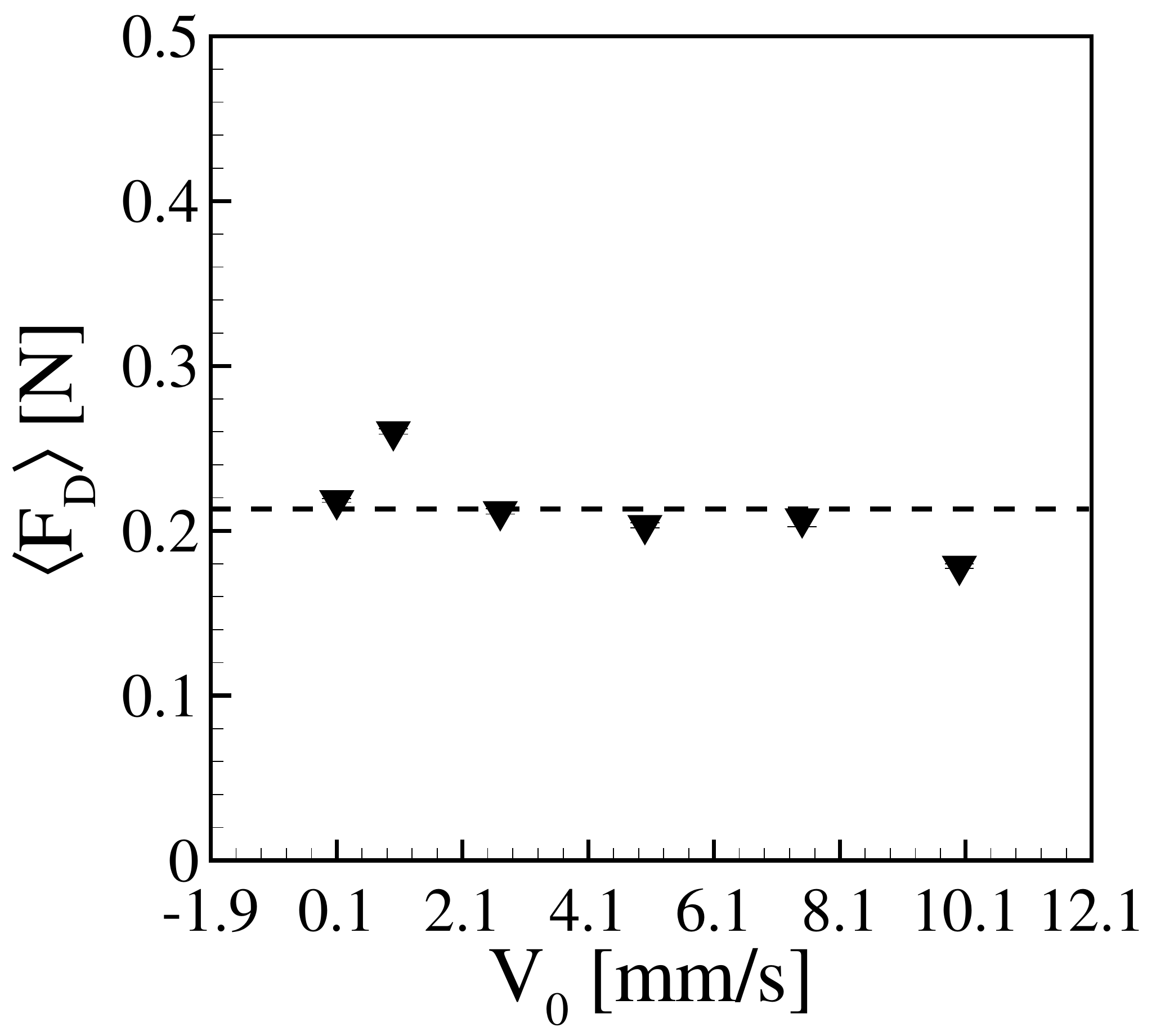}\\
				(b)
			\end{tabular}
		\end{minipage}
		\hfill
		\begin{minipage}{0.49\linewidth}
			\begin{tabular}{c}
				\includegraphics[width=0.80\linewidth]{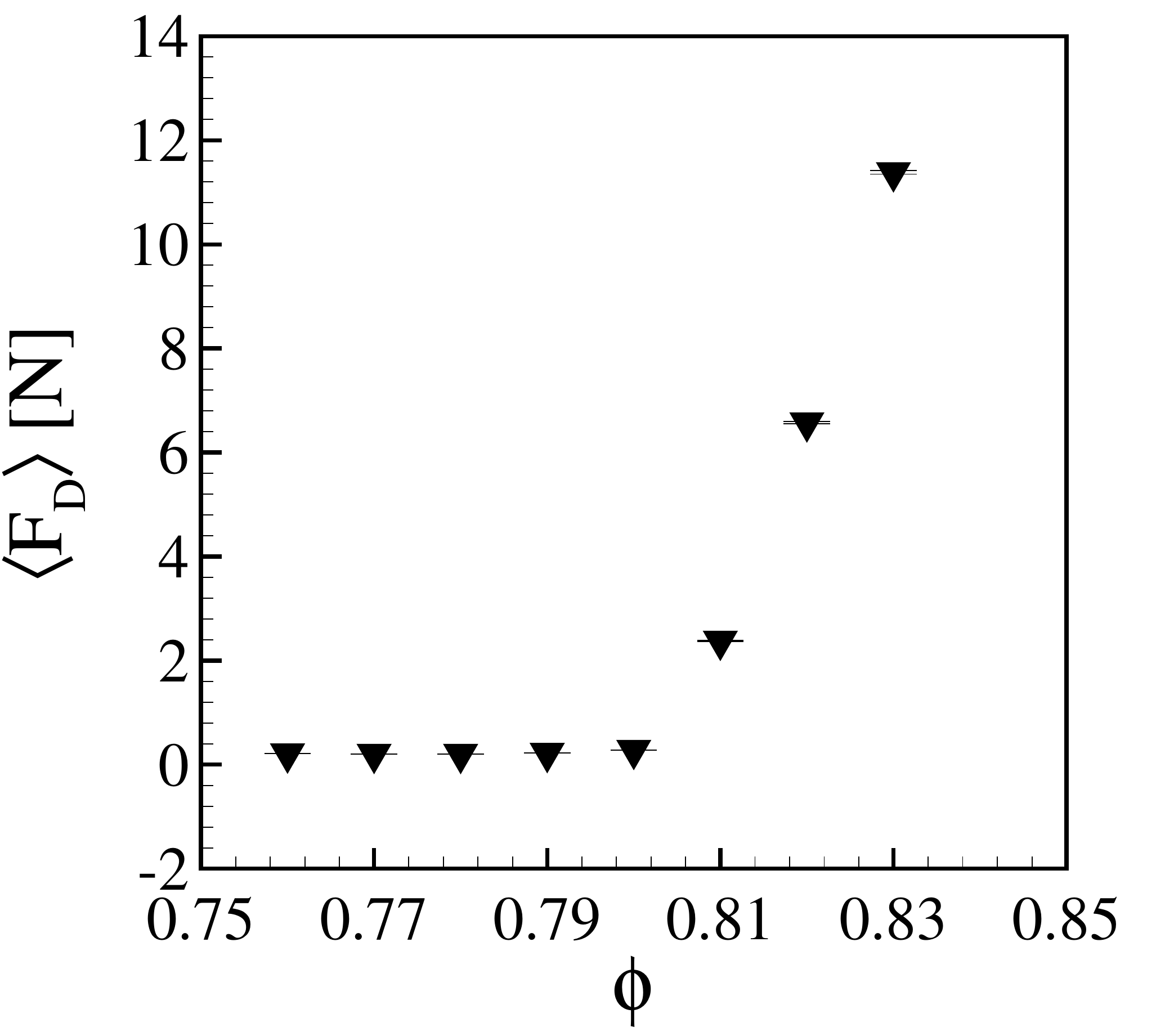}\\
				(c)
			\end{tabular}
		\end{minipage}
		\hfill
	\end{center}
	\caption{(a) Magnitude of the resultant force on the intruder $F_D$ when the mean packing fraction is $\phi$ = 0.76 and $V_0$ = 2.7 mm/s as a function of time $t$. (b) Time-averaged magnitude of the resultant force on the intruder $\left< F_D \right>$ as a function of its velocity for $\phi$ = 0.76. (c) $\left< F_D \right>$ as a function of $\phi$ for $V_0$ = 2.7 mm/s. In Figs. (b) and (c), symbols correspond to the average values and bars to the standard errors.}
	\label{fig:forces}
\end{figure}

For each simulated condition, we computed the resultant force on the intruder at each time step and associated it with the instantaneous drag force on the intruder $\vec{F}_D$. We then obtained its magnitude $F_D$ and, for each different condition, the time-averaged magnitude $\left< F_D \right>$. Figure \ref{fig:forces}a presents $F_D$ as a function of time $t$ when the intruder moves with $V_0$ = 2.7 mm/s in a system with mean packing fraction $\phi$ = 0.76. We observe an initial transient, when the intruder begins moving and $F_D$ increases due to an increasing number of contacts (shown next in Fig. \ref{fig:network_general}b), and that afterward the mean value remains roughly constant, with very high oscillations with peaks reaching values 3 times the mean value. Those strong oscillations are caused by the formation and destruction of contact networks that percolate forces within the bed, as shown next. The same behavior was found experimentally by Seguin et al. \cite{Seguin1}. Figure \ref{fig:forces}b shows the time-averaged magnitude of the drag force $\left< F_D \right>$ as a function of the intruder velocity $V_0$ for $\phi$ = 0.76. We obtain values that are roughly independent of $V_0$ (as in Refs. \cite{Seguin1} and \cite{Seguin3}), with a mean value $\left< F_D \right>$ $\approx$ 0.21. For the same case, Seguin et al. \cite{Seguin1} found experimentally $\left< F_D \right>$ $\approx$ 0.22. Finally, Fig. \ref{fig:forces}c presents $\left< F_D \right>$ as a function of $\phi$ for $V_0$ = 2.7 mm/s, showing that the mean force remains roughly constant until 0.80 $\leq$ $\phi$ $\leq$ 0.81, and from $\phi$ $\approx$ 0.81 on $\left< F_D \right>$ increases strongly with $\phi$, similar to results obtained experimentally by Kolb et al. \cite{Kolb1}. Considering that experimental uncertainties are expected in Refs. \cite{Kolb1, Seguin1} and that we obtained the particle properties (with the exception of the diameter) from other works, the agreement is good.

We observed also the formation of a cavity (void region) downstream the intruder, whose size decreases with increasing the packing fraction, as shown in Fig. \ref{fig:setup}d for 0.76 $\leq$ $\phi$ $\leq$ 0.83. We observe that for $\phi$ = 0.80 the cavity has almost disappeared, and for $\phi$ = 0.83 it no longer exists, in accordance with the experimental observations of Kolb et al. \cite{Kolb1}.

Altogether, the resultant drag and cavity agree with experimental observations and validate part of our numerical results. More information on the numerical simulations (input and output files, numerical scripts for post-processing the outputs, etc.) are available on a public repository \cite{Supplemental2}.

\subsection{\label{sec:networks} Network of contact forces}

In the following, we analyze the network of contact forces and the behavior of individual grains within specific contact chains. For that, we fixed the mean particle fraction to $\phi$ = 0.76 and the intruder velocity to $V_0$ = 2.7 mm/s. From images of the force chains, such as Fig. \ref{fig:setup}c, we observe that forces from the intruder propagate through contact networks whose anisotropy and size depend on the force level and region within the system. In what follows, we investigate the anisotropy of the system (i) as a whole, (ii) for different force levels (below and above an average value), and (iii) for different regions within the domain. For that, we computed the fabric tensor $\hat{R}$ \cite{Bi},

\begin{equation}
	\hat{R} = \frac{1}{N}\sum_{i \neq j}\frac{\mathbf{r}_{ij}}{|\mathbf{r}_{ij}|}\otimes\frac{\mathbf{r}_{ij}}{|\mathbf{r}_{ij}|},
	\label{eqnfabricR}
\end{equation}

\noindent where $N$ is the number of non-rattler particles (particles with at least two contacts), $\textbf{r}_{ij}$ is the contact vector from the center of particle $i$ to the contact between particles $i$ and $j$, and $\otimes$ denotes the vector outer product. With the eigenvalues $R_{1}$ and $R_{2}$ of the tensor $\hat{R}$, we computed the average number of contacts per particle $Z = R_{1} + R_{2}$ and the anisotropy of the contact network $\rho = R_{1} - R_{2}$ \cite{Bi}.

\begin{figure}[h!]
	\begin{center}
		\begin{minipage}{0.49\linewidth}
			\begin{tabular}{c}
				\includegraphics[width=0.80\linewidth]{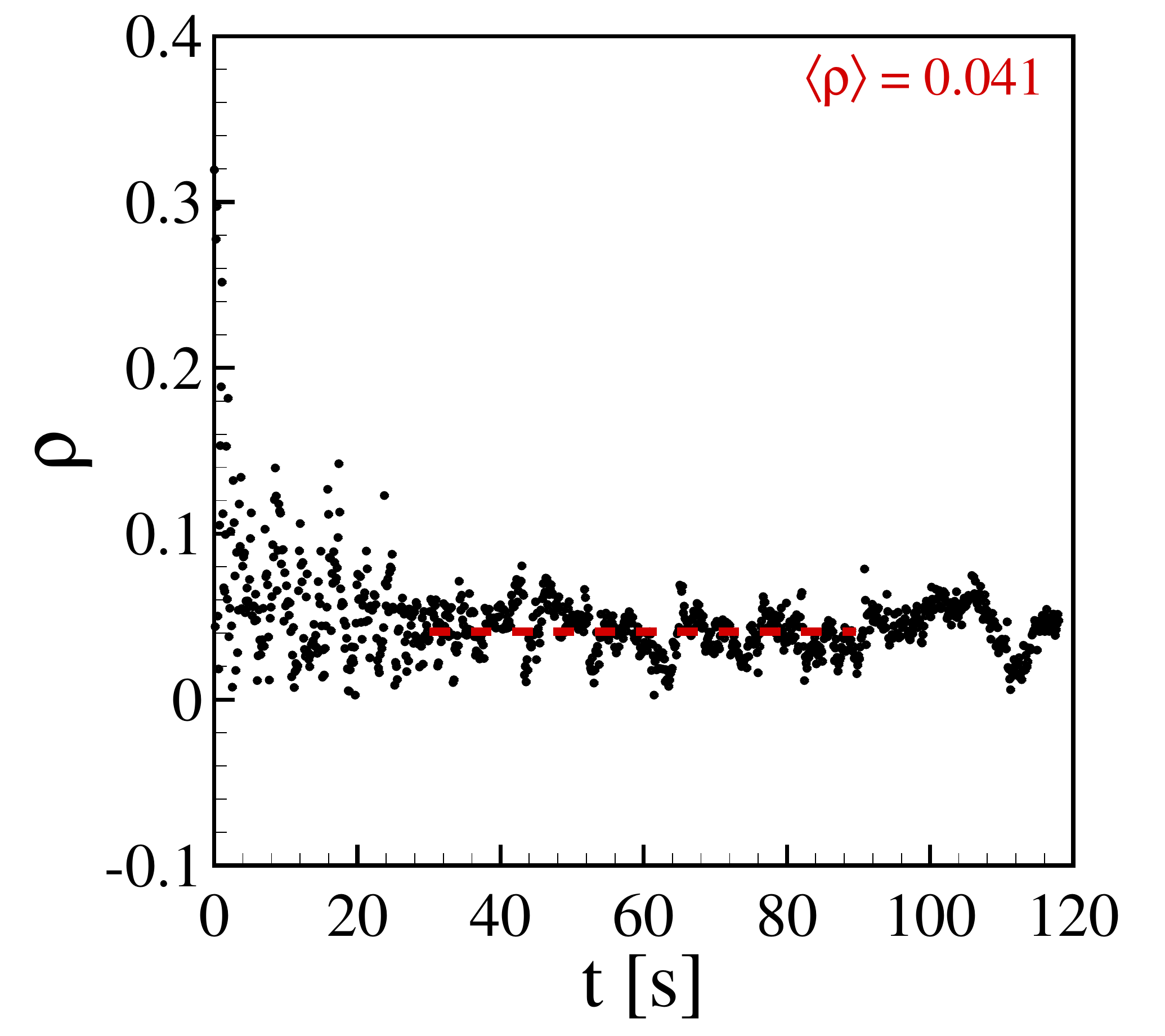}\\
				(a)
			\end{tabular}
		\end{minipage}
		\hfill
		\begin{minipage}{0.49\linewidth}
			\begin{tabular}{c}
				\includegraphics[width=0.80\linewidth]{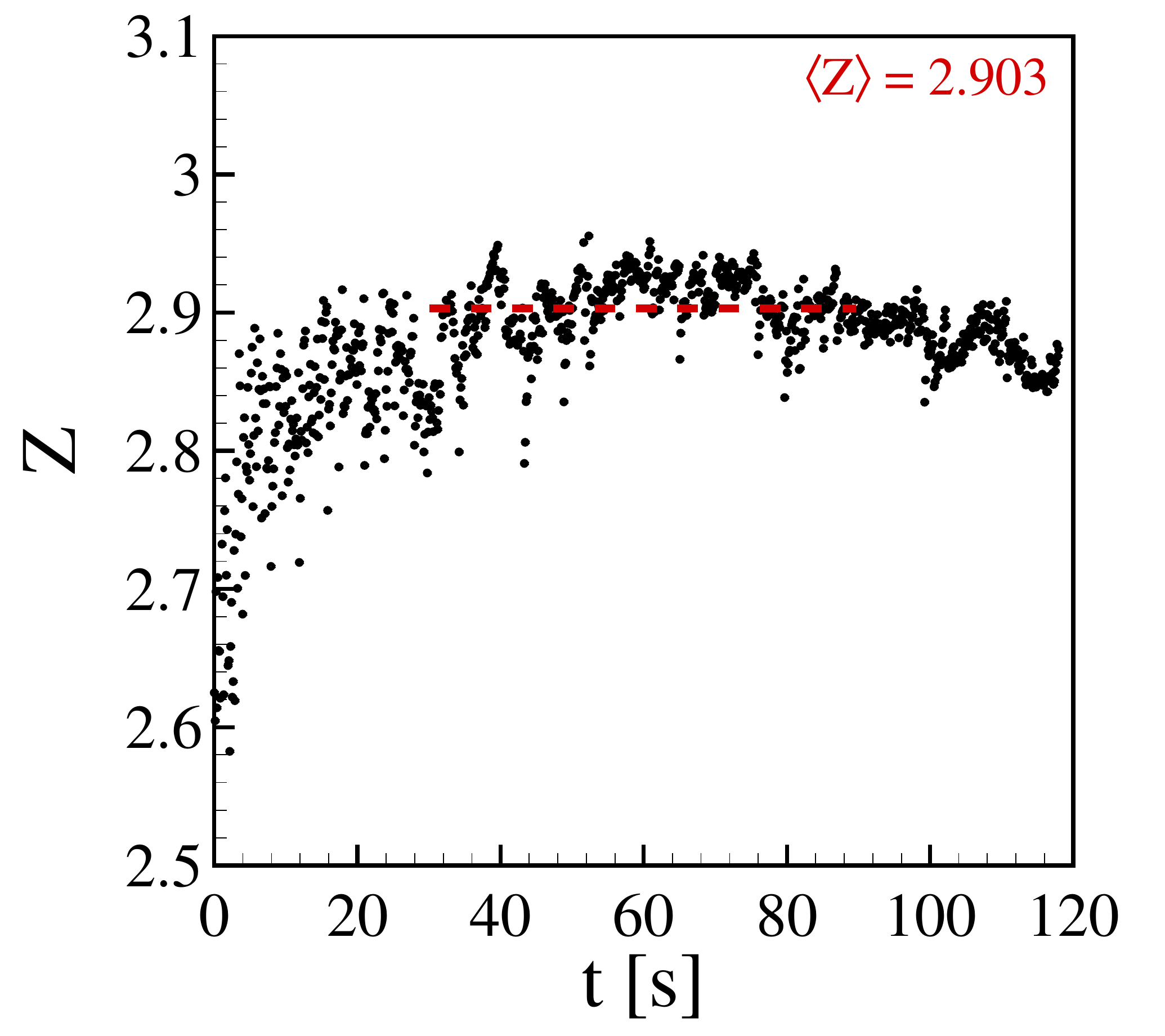}\\
				(b)
			\end{tabular}
		\end{minipage}
		\hfill
	\end{center}
	\caption{Time evolution of (a) contact anisotropy $\rho$ and (b) average number of contacts per particle $Z$ for the entire system.}
	\label{fig:network_general}
\end{figure}

Figures \ref{fig:network_general}a and \ref{fig:network_general}b present the contact anisotropy $\rho$ and average number of contacts per particle $Z$, respectively, for the entire domain as functions of time (the time evolution of the number of non-rattler particles $N$ is available in the Supplemental Material \cite{Supplemental}). We observe that $Z$ increases and $\rho$ decreases during the first 20 s. The strong initial variations of $Z$ and $\rho$ are mainly due to adaptations of the initial conditions of the system as the intruder starts moving, with more grains being put into contact and a general decrease in anisotropy. After this time interval, mean values present lower variations. At the stable intervals ($t$ $>$ 30 s), time averages computed for the ROI are $\left< \rho \right>$ = 0.041 and $\left< Z \right>$ = 2.903, showing that, when considered as a whole in terms of regions and force magnitudes, the contact network has a low degree of anisotropy. However, since the intruder moves in one direction, we expect load-bearing chains aligned in preferential directions in order to resist to the intruder's motion \cite{Cates, Majmudar, Bi}. Load-bearing chains that transmit strong forces have been shown to exist in compressed 2D granular systems \cite{Radjai1, Seguin2}, to be more anisotropic than the dissipative chains, and, in addition, to be related to jamming by shear \cite{Bi}. We investigate next if this is also the case for a 2D system displaced by an intruder, and, in addition, if anisotropy varies in space.

\subsubsection{\label{sec:force_levels} Force levels}

Following the same idea of Radjai et al. \cite{Radjai1}, we divided the network of contact forces into a bearing network, for which transmitted forces are higher than the average value (ensemble average at each considered instant), and a dissipative network, with values lower than the average. Once identified the type of network, we computed $\hat{R}$, $\rho$ and $Z$ for each network (all chains), which are shown in this subsection. We also followed the evolution of specific chains and the motion of their grains, which are shown in Subsection \ref{sec:bearing_chain}.

\begin{figure}[h!]
	\begin{center}
		\begin{minipage}{0.49\linewidth}
			\begin{tabular}{c}
				\includegraphics[width=0.80\linewidth]{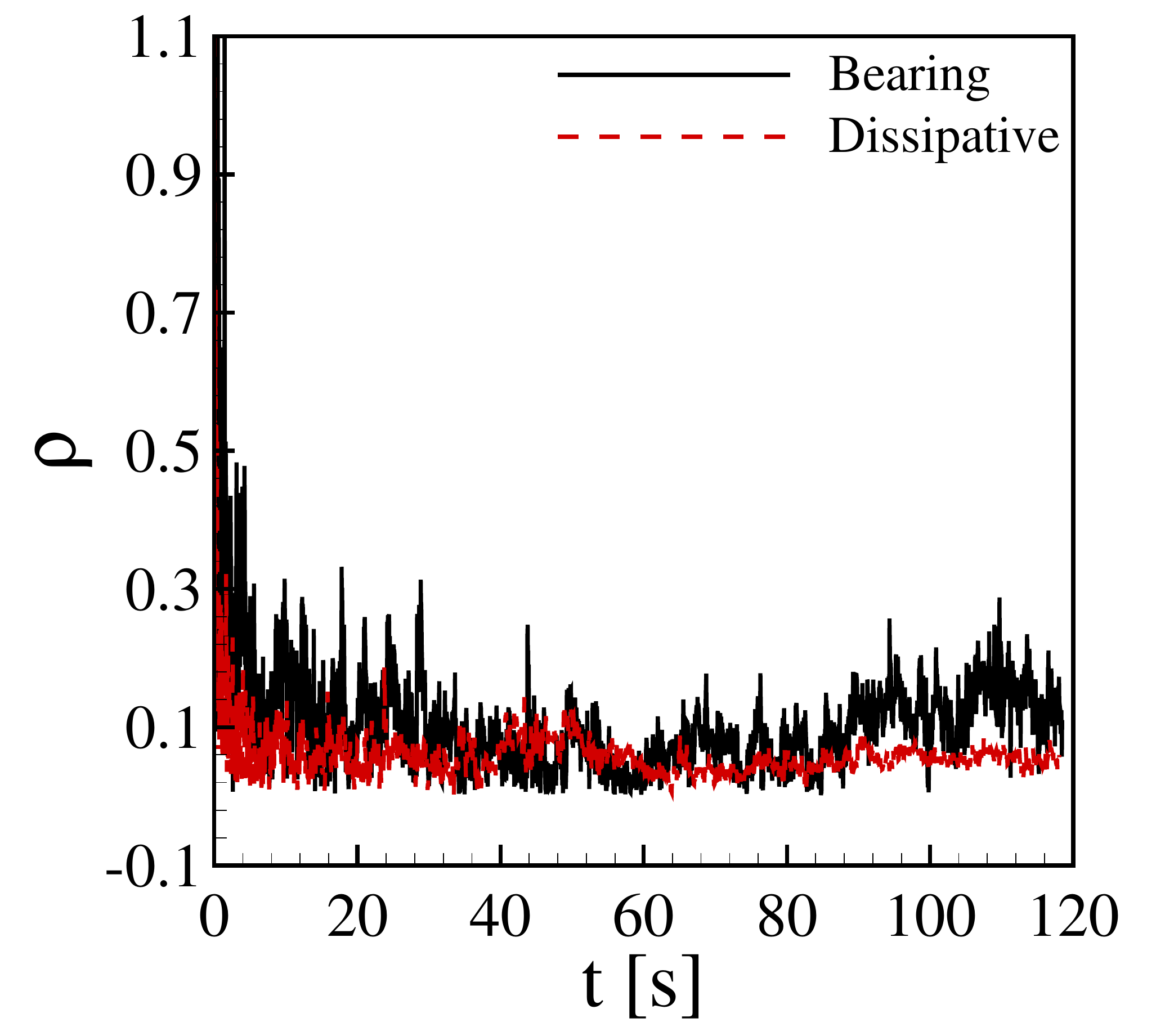}\\
				(a)
			\end{tabular}
		\end{minipage}
		\hfill
		\begin{minipage}{0.49\linewidth}
			\begin{tabular}{c}
				\includegraphics[width=0.80\linewidth]{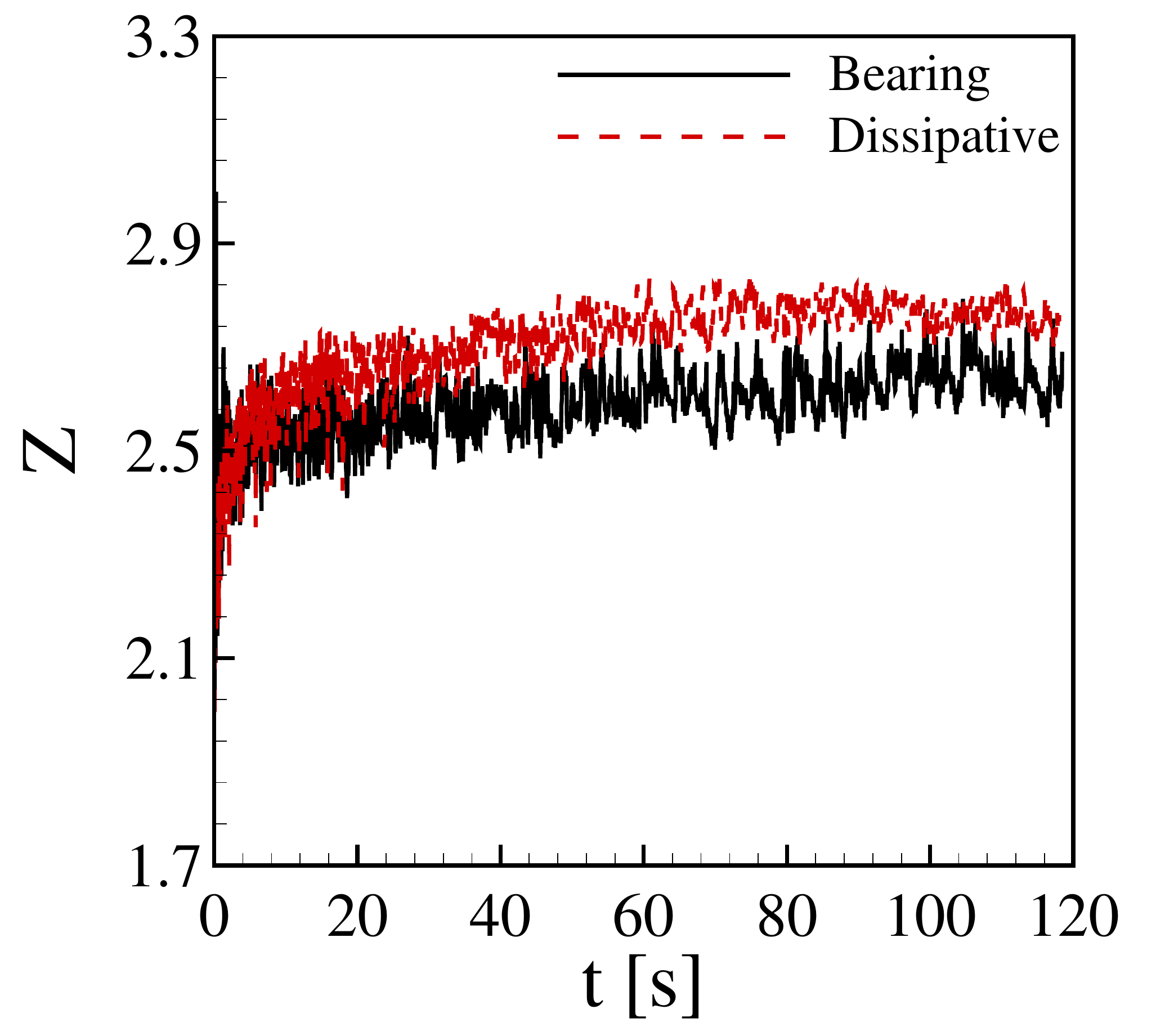}\\
				(b)
			\end{tabular}
		\end{minipage}
		\hfill
	\end{center}
	\caption{Time evolution of (a) contact anisotropy $\rho$ and (b) average number of contacts per particle $Z$, computed for the bearing (continuous line) and dissipative (dashed-red line) networks. Time average values are $\left< \rho \right>$ = 0.107 and $\left< Z \right>$ = 2.592 for the bearing network, and $\left< \rho \right>$ = 0.060 and $\left< Z \right>$ = 2.711 for the dissipative network.}
	\label{fig:network_divided}
\end{figure}

Figures \ref{fig:network_divided}a and \ref{fig:network_divided}b present the time evolution of the contact anisotropy $\rho$ and average number of contacts per particle $Z$, respectively, for the bearing (strong) and dissipative (weak) networks for the entire domain (the time evolution of $N$ is available in the Supplemental Material \cite{Supplemental}). We observe an increase in $Z$ during the first 20 s for the dissipative network while for the bearing network the mean value of $Z$ remains roughly constant, with values for the dissipative network being 5 \% higher than those for the bearing network. During the first 10-20 s for both networks, $\rho$ decreases, with values 80 \% higher for the bearing network in comparison with the dissipative network. These values indicate that anisotropy is maintained mostly by the load-bearing chains. From direct observations of figures of the network of contact forces, such as Figs. \ref{fig:setup}c and \ref{fig:network_basal_reduction}, or from the animation available in the Supplemental Material \cite{Supplemental}, we observe that bearing chains percolate in various directions, but mostly in the front (upstream) region of the intruder. This characteristic, which is similar to the shear jammed state described by Bi et al \cite{Bi} for the case of a sheared cell, explains the higher anisotropy of the bearing network. This is also in accordance with the description given by Kolb et al. \cite{Kolb1} of a jammed region in front of the intruder, with load-bearing chains being formed and collapsing as the intruder moves, making the drag force to fluctuate strongly around a mean value, as shown in Fig. \ref{fig:forces}a. We investigate the bearing chains in detail (at the grain scale) in Subsection \ref{sec:bearing_chain}.

\subsubsection{\label{sec:spatial_distr} Spatial distribution}

\begin{figure}[h!]
	\begin{center}
		\includegraphics[width=0.95\columnwidth]{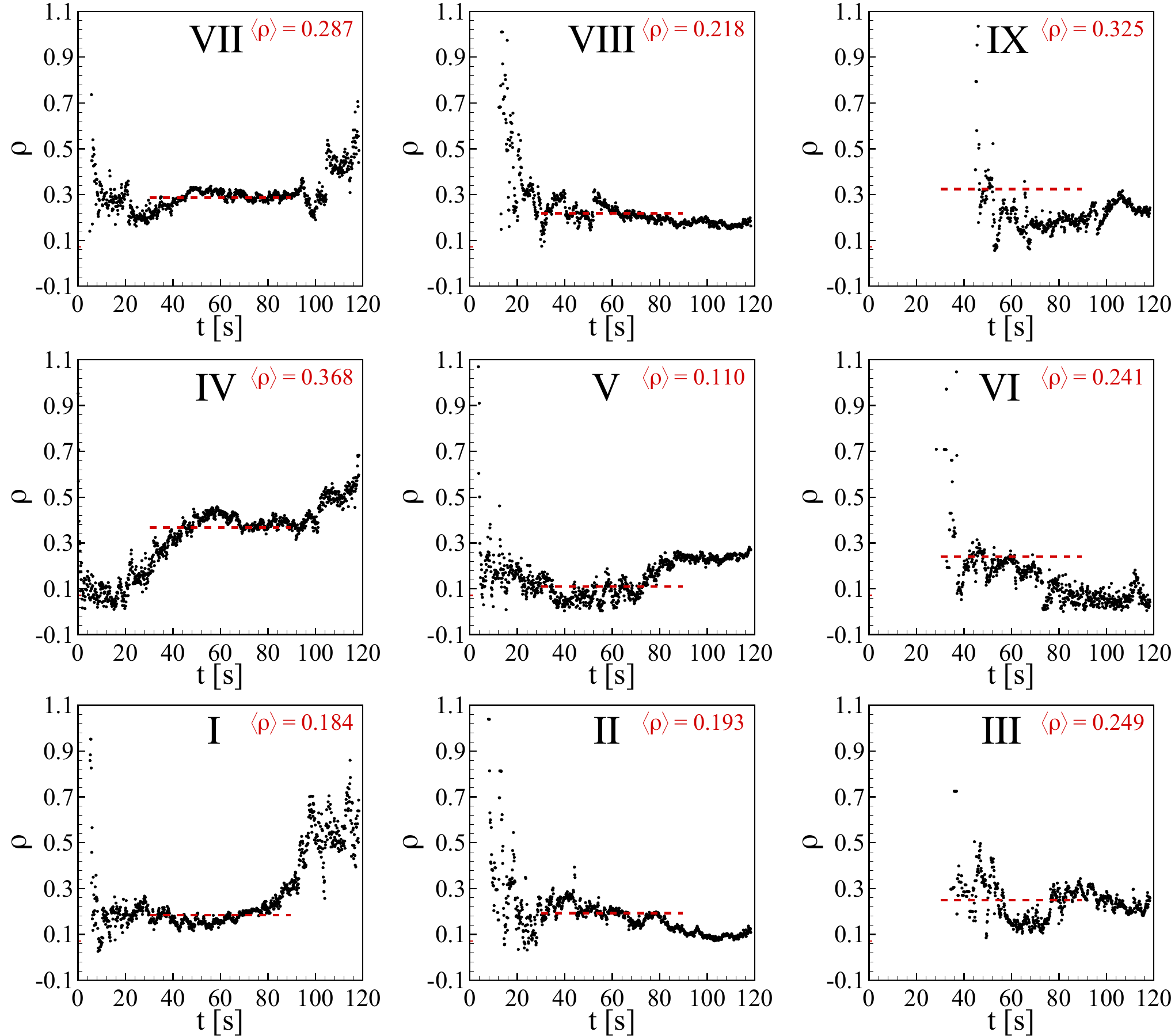}\\ 
	\end{center}
	\caption{Time evolution of the anisotropy of the contact network $\rho$ for each individual region of Fig. \ref{fig:setup}b.}
	\label{fig:rho_regions}
\end{figure}

\begin{figure}[h!]
	\begin{center}
		\includegraphics[width=0.95\columnwidth]{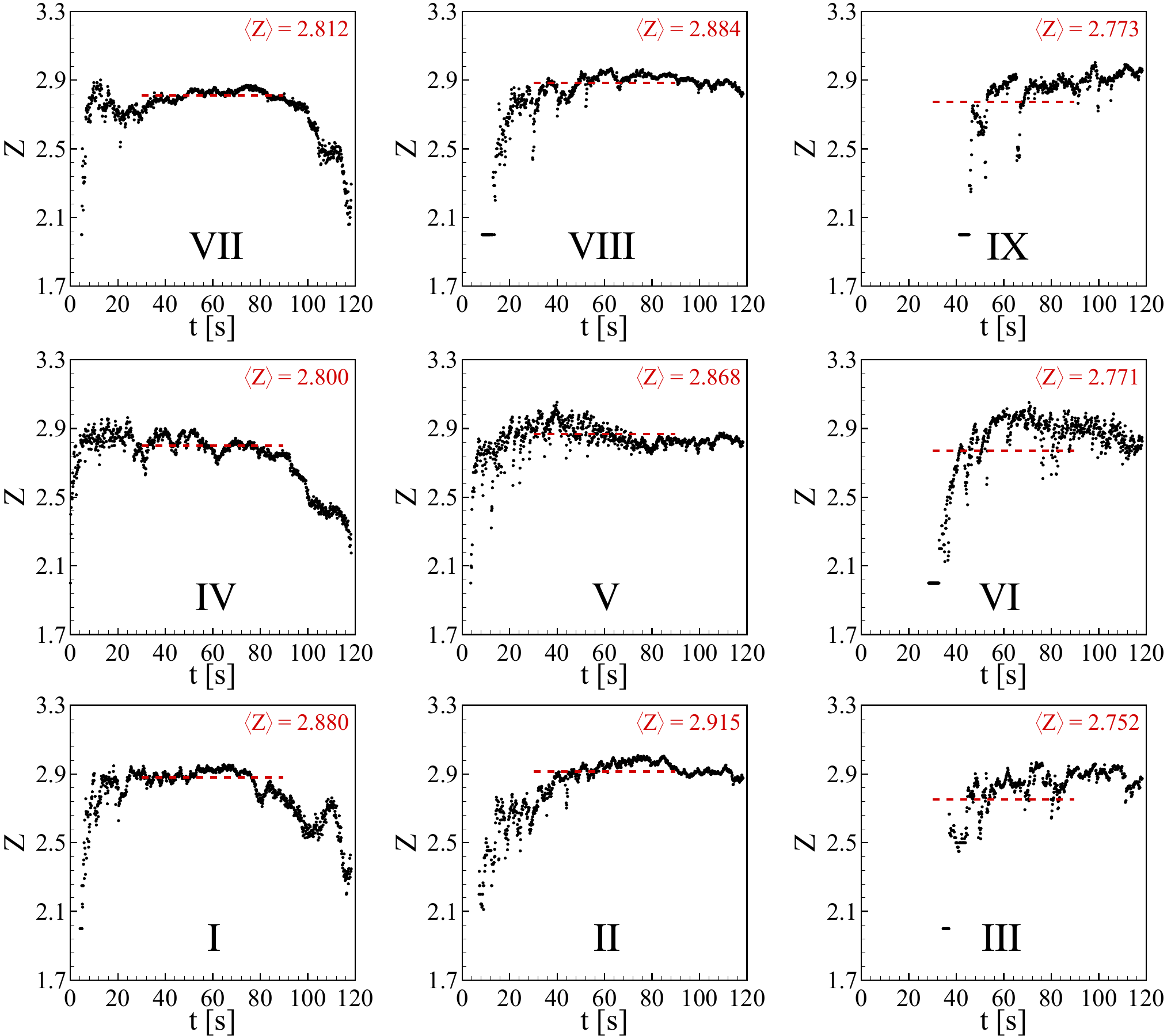}\\ 
	\end{center}
	\caption{Time evolution of the average number of contacts per particle $Z$ for each individual region of Fig. \ref{fig:setup}b.}
	\label{fig:z_regions}
\end{figure}

In order to investigate if the time evolution of anisotropy varies in space, we divided the domain in nine different regions as shown in Fig. \ref{fig:setup}b. Figure \ref{fig:rho_regions} presents the anisotropy of the contact network $\rho$ for each individual region of Fig. \ref{fig:setup}b as a function of time (the relative positions of graphics correspond to the spatial distribution in the domain). Interestingly, $\rho$ decreases in regions mainly upstream the intruder (regions II, III, V, VI, VIII and IX) as the latter moves from the left to the right (Fig. \ref{fig:setup}), while $\rho$ increases in the regions farther downstream the intruder (regions I, IV and VII) as it approaches the right boundary of the domain. In addition, values of $\rho$ are much higher in the left regions by the end of the intruder's motion (values at least three times greater on the left than on the right regions). This behavior is corroborated by the time evolution of $Z$, shown in Fig. \ref{fig:z_regions} for each region. From this figure, we observe that the average number of contacts per particle decreases in the left regions (regions I, IV, and VII) by the end of motion, compatible with anisotropic behaviors, while the same does not occur in the other regions.

The explanation for the long-range effects is the size of the contact network that, by the end of motion of the intruder, reaches regions far downstream of it (see the Supplemental Material \cite{Supplemental} for an animation showing the instantaneous contact network). Because chains arriving at the farther regions (I, IV and VII) by the end of the intruder's motion follow principal directions, as if irradiating from the intruder, anisotropy is larger in those regions. These results corroborate the necessity of a non-local rheology to describe a granular system displaced by an intruder, even if most of grain displacements occur in the vicinity of the intruder (as shown in Refs. \cite{Kolb1, Seguin1} and in Subsection \ref{sec:bearing_chain}).

\subsubsection{\label{sec:bearing_chain} Grains within bearing chains}

\begin{figure}[h!]
	\begin{center}
		\begin{tabular}{c}
			\includegraphics[width=0.45\columnwidth]{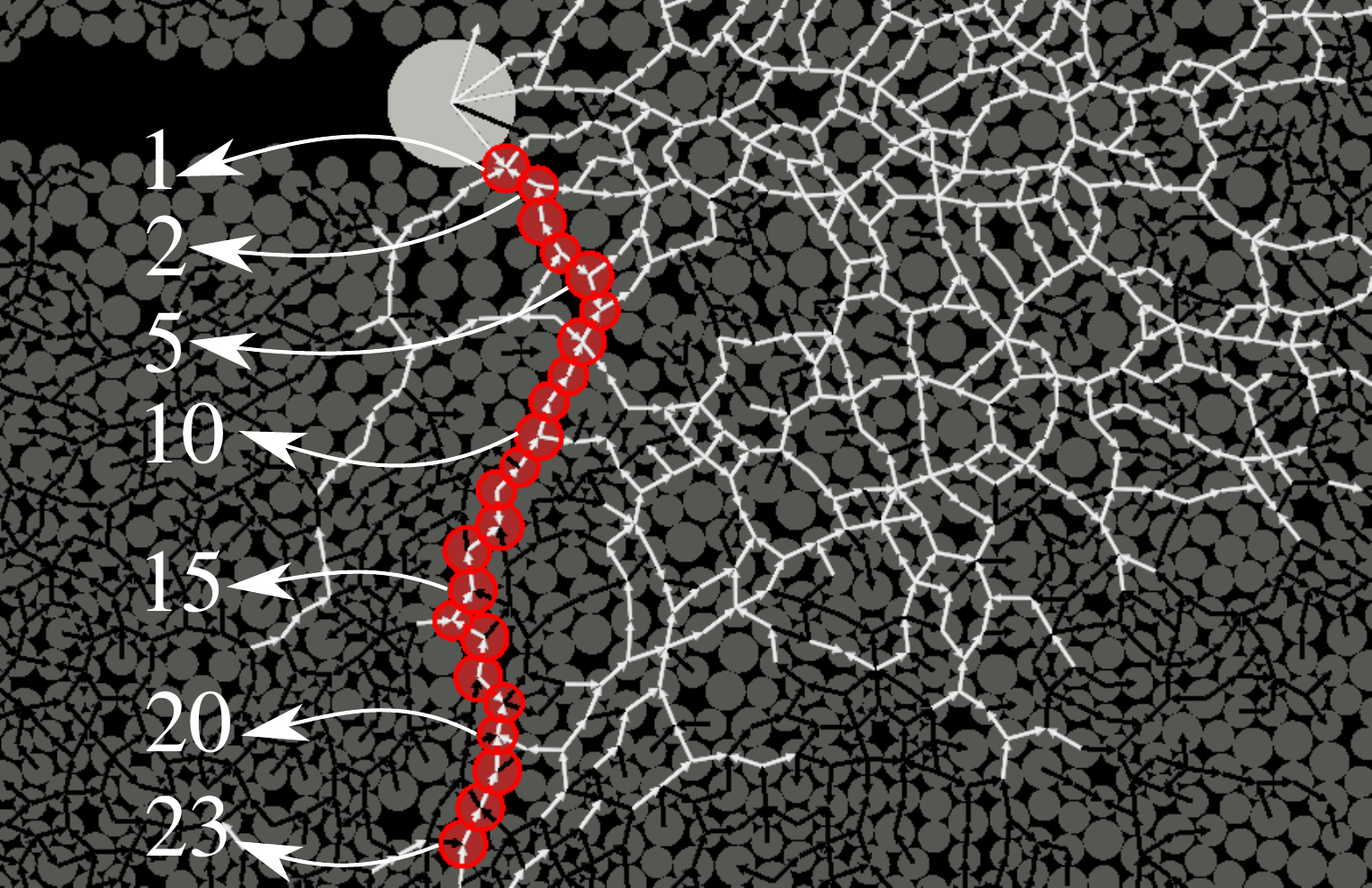}\\
			(a)\\
			\\
			\includegraphics[width=0.45\columnwidth]{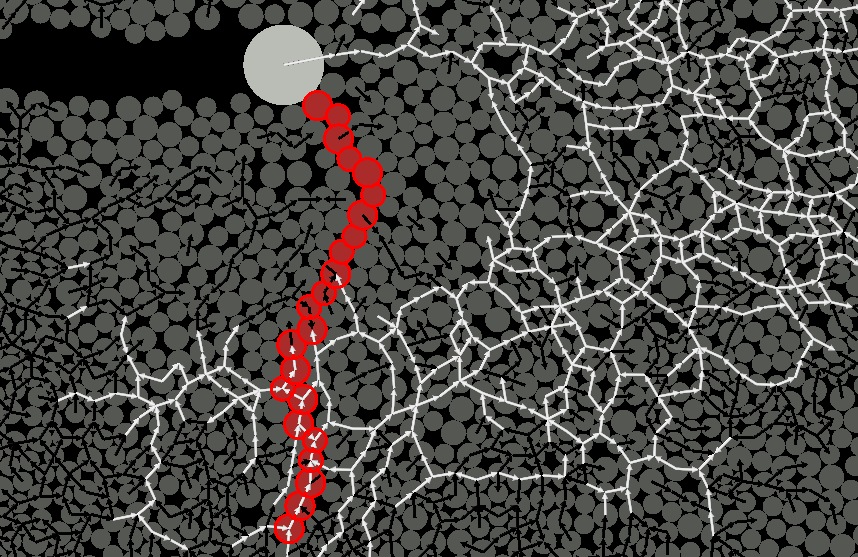}\\
			(b)
		\end{tabular} 
	\end{center}
	\caption{Grains within a bearing chain (shown in red) of the network shown in Fig. \ref{fig:setup}c (bearing network in white). (a) Grains just after the chain was formed, with labels from 1 to 23 according to their distance from the intruder. (b) Grains just after the considered chain broke.}
	\label{fig:chain_bearing_grains}
\end{figure}

The experimental results of Kolb et al. \cite{Kolb1} and Seguin et al. \cite{Seguin1} showed strong fluctuations of $F_D$ around a mean value that are associated  with the formation and breaking of bearing chains, and the same behavior appears in our simulations. However, previous works did not show how grains within a bearing chain move nor how the chain breaks. We investigate this problem by choosing some bearing chains, labeling the grains of each chain, and following these grains along time. For the labels, the corresponding numbers start with the grain in contact with the intruder and increase as grains are farther from it, as shown in Fig. \ref{fig:chain_bearing_grains}a.

One example of bearing chain is shown in Fig. \ref{fig:chain_bearing_grains}, where Fig. \ref{fig:chain_bearing_grains}a shows the grains just after the chain was formed and Fig. \ref{fig:chain_bearing_grains}b after the chain broke. We observe that during this period some chains broke while some others formed, and that the motions of the considered grains are very small. In order to inquire into the motions of these grains, we computed their fluctuations with respect to the ensemble of grains and their accumulated displacements, and found that chains break due to creep motion of some grains, with a very small degree of fluctuations of individual grains. Because the oscillation levels of load-bearing grains are negligible, we present next only their displacements. 

\begin{figure}[h!]
		\begin{center}
		\begin{minipage}{0.49\linewidth}
			\begin{tabular}{c}
				\includegraphics[width=0.80\linewidth]{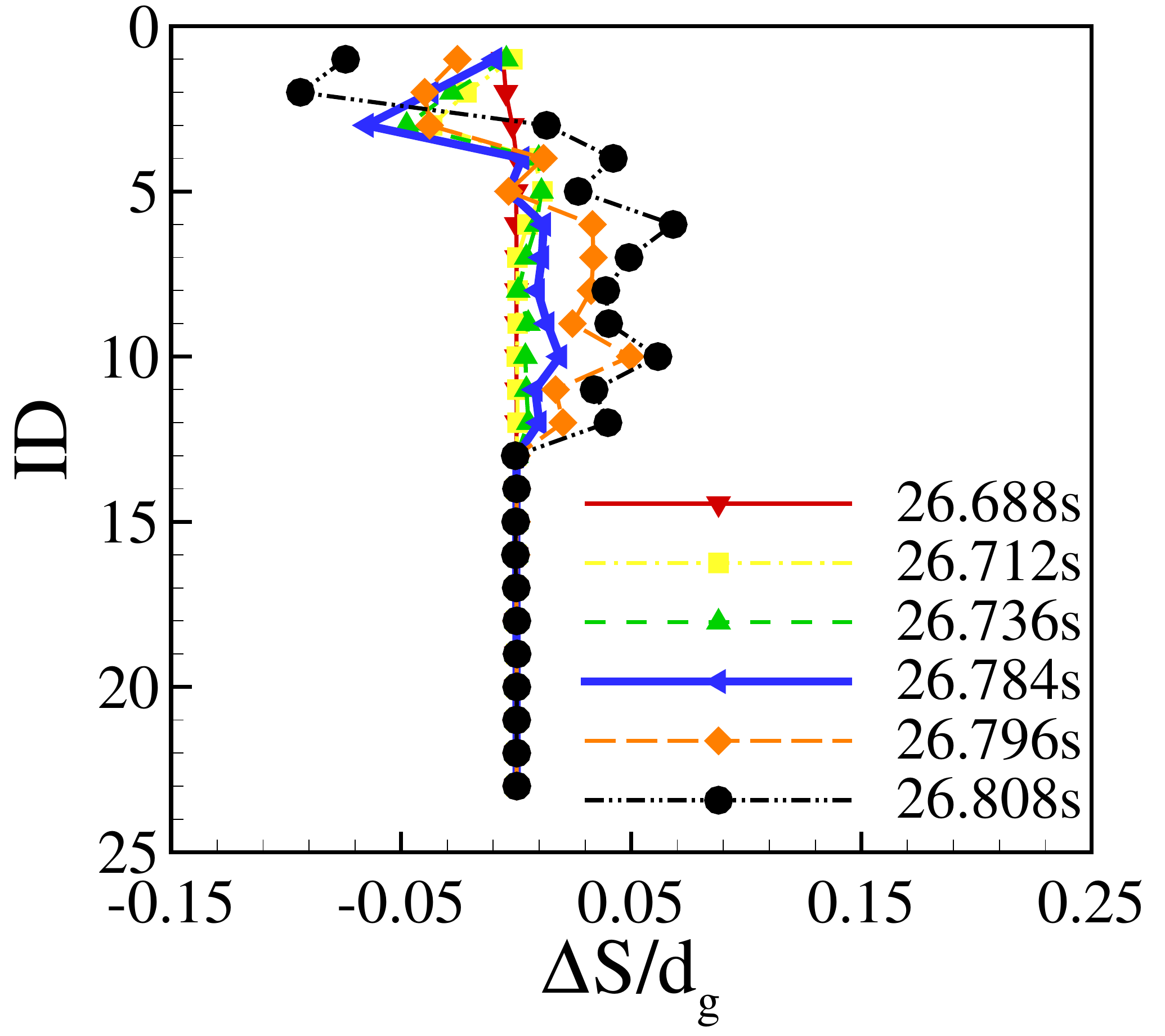}\\
				(a)
			\end{tabular}
		\end{minipage}
		\hfill
		\begin{minipage}{0.49\linewidth}
			\begin{tabular}{c}
				\includegraphics[width=0.80\linewidth]{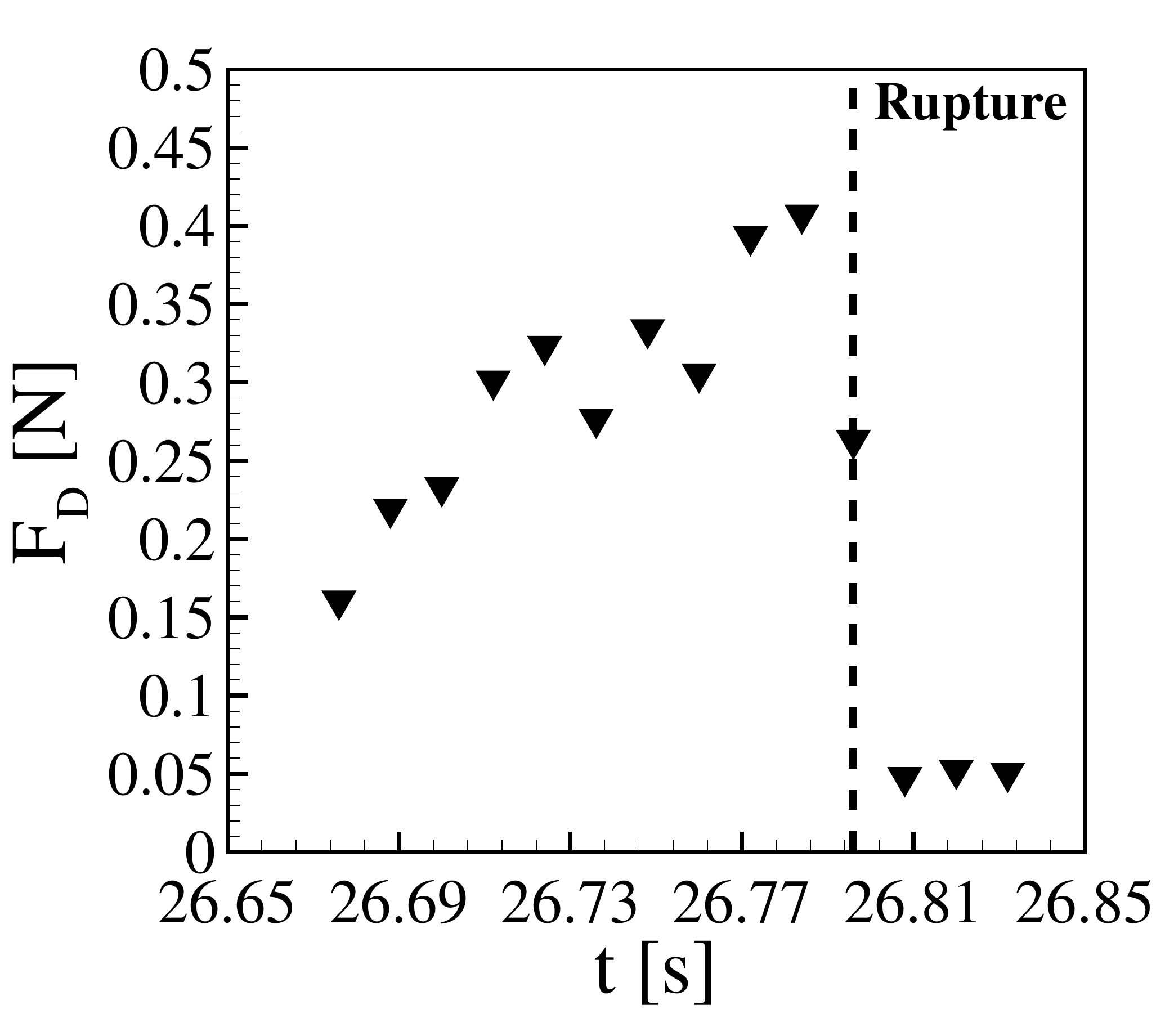}\\
				(b)
			\end{tabular}
		\end{minipage}
		\hfill
		\begin{minipage}{0.49\linewidth}
			\begin{tabular}{c}
				\includegraphics[width=0.80\linewidth]{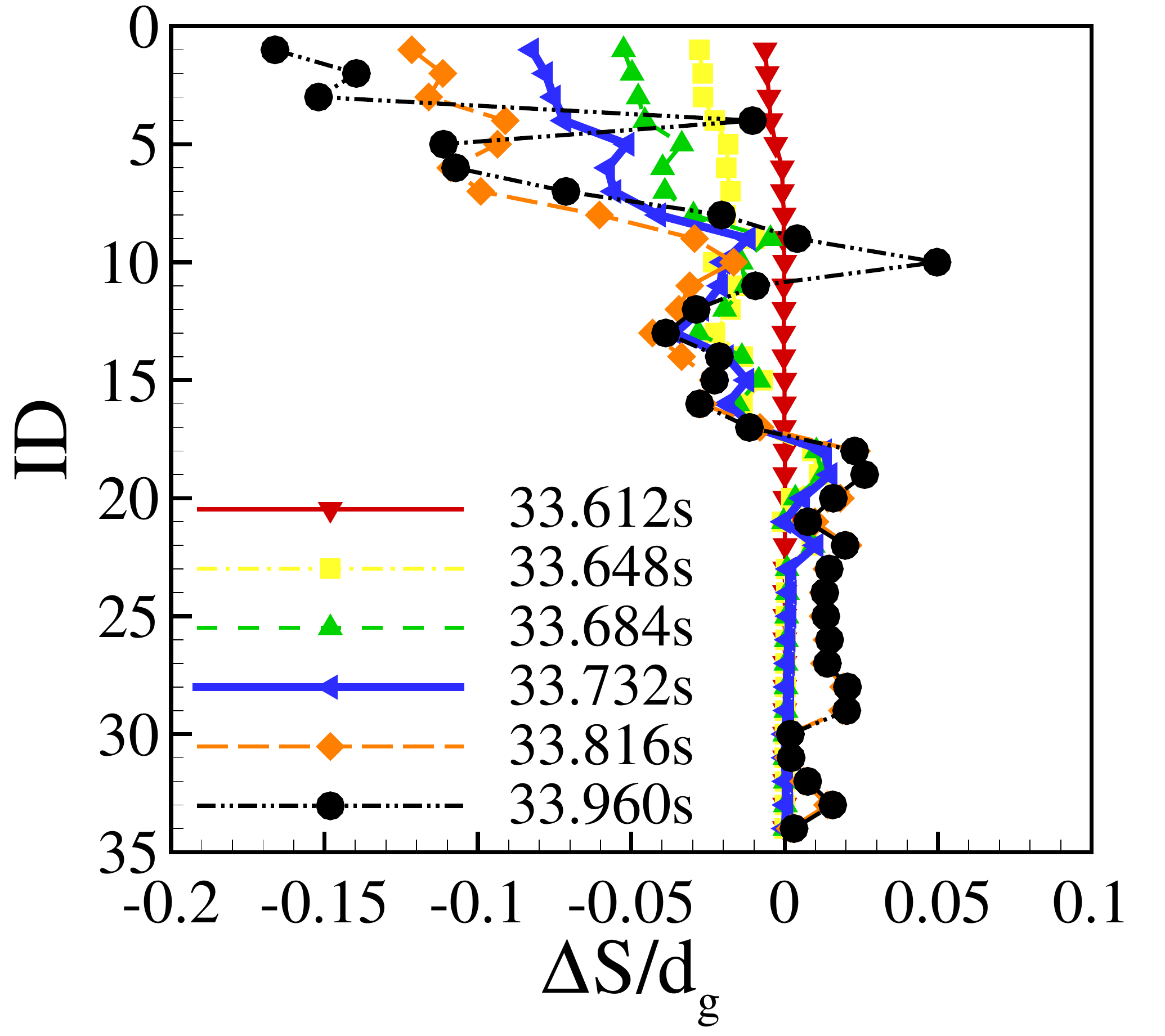}\\
				(c)
			\end{tabular}
		\end{minipage}
		\hfill
		\begin{minipage}{0.49\linewidth}
			\begin{tabular}{c}
				\includegraphics[width=0.80\linewidth]{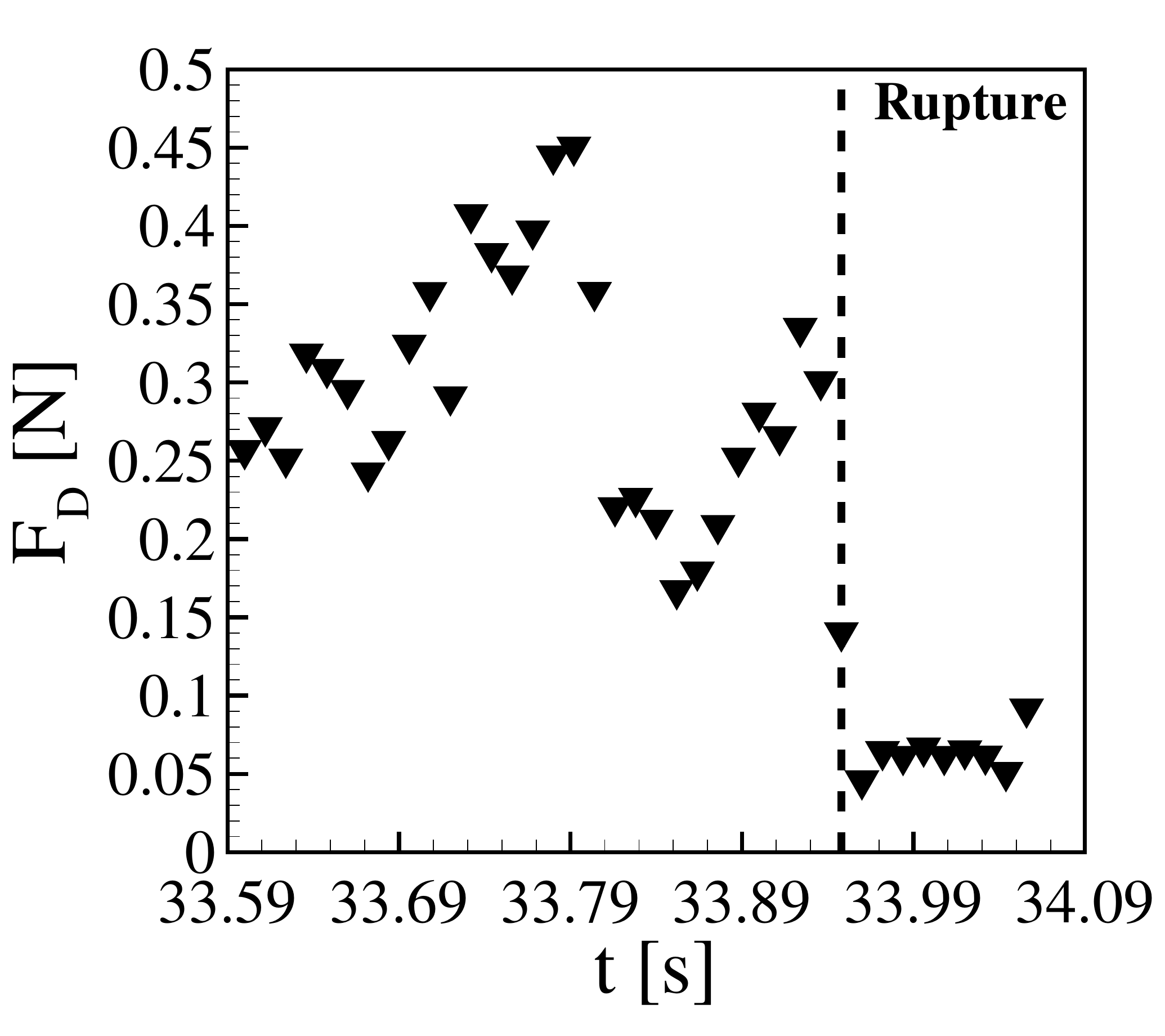}\\
				(d)
			\end{tabular}
		\end{minipage}
		\hfill
	\end{center}
	\caption{Displacements of each labeled grain for different instants during the breaking of a bearing chain (a) transverse and (c) parallel to the intruder's motion, normalized by the mean grain diameter $d_g$ = 4.5 mm. Figure (a) corresponds to the same chain shown in Fig. \ref{fig:chain_bearing_grains}. Magnitude of the drag force on the intruder $F_D$ along time during the formation and breaking of the (b) transverse and (d) longitudinal bearing chains. The dashed line indicates the instant when the chain is present for the last time, i.e., just before its rupture.}
	\label{fig:chain_creep}
\end{figure}

\begin{figure}[h!]
	\begin{center}
		\includegraphics[width=0.45\columnwidth]{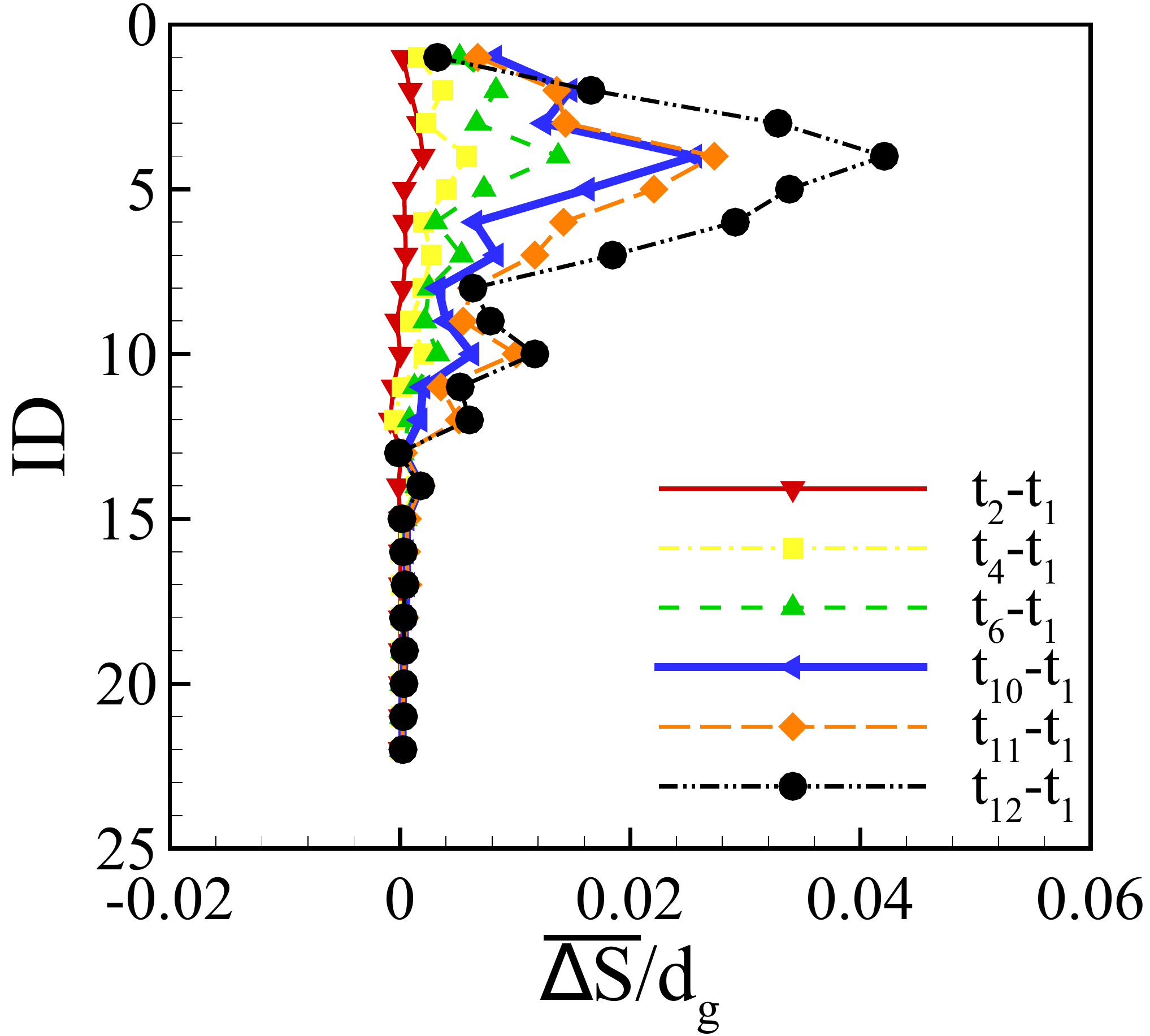}\\ 
	\end{center}
	\caption{Ensemble average of the displacements of grains $\overline{\Delta S}$ for different instants during the breaking of bearing chains transverse to the intruder's motion, normalized by the mean grain diameter $d_g$.}
	\label{fig:chain_creep_average}
\end{figure}

Figure \ref{fig:chain_creep}a presents the displacements $\Delta S$ of each labeled grain at different instants for the chain shown in Fig. \ref{fig:chain_bearing_grains} (transverse to the intruder's motion), and Fig. \ref{fig:chain_creep}c for a chain parallel to the intruder's motion. Each symbol corresponds to one instant, and the figures represent the strain suffered by the chain. We observe that while the intruder is forced through the system the grains closer to it yield and move, while those farther (labeled 14 or more in Fig. \ref{fig:chain_creep}a and 19 or more in Fig. \ref{fig:chain_creep}c) do not move. In fact, we can observe from Fig. \ref{fig:chain_bearing_grains}b that the latter remain in contact with each other, so that the part of the chain that is not in contact with the intruder persists. The same behavior was observed for all the chains that we tracked. Figure \ref{fig:chain_creep_average} presents an ensemble average computed for 11 chains, for which we note that, in average, creep during chain breaking is localized around the intruder, decreasing fast as grains are farther from the intruder and being nonexistent from the 15$^{th}$ grain on. The average duration of creeping $\Delta t$ is of the order of 0.1 s, and normalizing $\Delta t$ by the characteristic time $t_c$ = $d_g / V_0$, where $d_g$ = 4.5 mm is the mean grain diameter, we obtain $\Delta t / t_c$ of the order of 0.1 (see the Supplemental Material \cite{Supplemental} for a table showing the duration of each chain). In terms of drag on the intruder, Figs. \ref{fig:chain_creep}b and \ref{fig:chain_creep}d show the time evolution of $F_D$ as the same bearing chains of Figs. \ref{fig:chain_creep}a and \ref{fig:chain_creep}c, respectively, break. We observe the same behavior shown by Kolb et al. \cite{Kolb1}: an increase in $F_D$ while the bearing chain persists, and a fast decrease when the chain breaks.

In summary, we observe that the formation and breaking of bearing chains are responsible for the high oscillation levels of $F_D$, with $F_D$ increasing during the lifetime of the bearing chain as the intruder is forced against it and decreasing as the chain breaks. In addition, we observe that the breaking occurs due to creeping of grains closer to the intruder while those farther from it do not move, and that, once broken, part of the former chain persists.

\subsection{\label{sec:basal_friction} Basal friction}

\begin{table}[h!]
	\caption{Coefficients of basal friction: values used in the simulations.}
	\label{tabcoefficients_reduction}
	\begin{tabular}{c|ccccccc}
		\hline
		 Coefficients & \multicolumn{7}{c}{Relative reductions}                                                                                                                                       \\ \hline
		 & \multicolumn{1}{c|}{100$\%$} & \multicolumn{1}{c|}{75$\%$} & \multicolumn{1}{c|}{50$\%$} & \multicolumn{1}{c|}{25$\%$} & \multicolumn{1}{c|}{10$\%$} & \multicolumn{1}{c|}{1$\%$} & 0.1$\%$ \\ \cline{2-8} 
		$\mu_{iw}$   & \multicolumn{1}{c|}{0.7}     & \multicolumn{1}{c|}{0.525}  & \multicolumn{1}{c|}{0.35}   & \multicolumn{1}{c|}{0.175}  & \multicolumn{1}{c|}{0.07}   & \multicolumn{1}{c|}{0.007} & 0.0007  \\
		$\mu_{gw}$   & \multicolumn{1}{c|}{0.4}     & \multicolumn{1}{c|}{0.3}    & \multicolumn{1}{c|}{0.2}    & \multicolumn{1}{c|}{0.1}    & \multicolumn{1}{c|}{0.04}   & \multicolumn{1}{c|}{0.004} & 0.0004  \\
		$\mu_{s,gw}$ & \multicolumn{1}{c|}{0.7}     & \multicolumn{1}{c|}{0.525}  & \multicolumn{1}{c|}{0.35}   & \multicolumn{1}{c|}{0.175}  & \multicolumn{1}{c|}{0.07}   & \multicolumn{1}{c|}{0.007} & 0.0007  \\ \hline
	\end{tabular}
\end{table}

\begin{figure}[h!]
	\begin{center}
		\includegraphics[width=0.70\columnwidth]{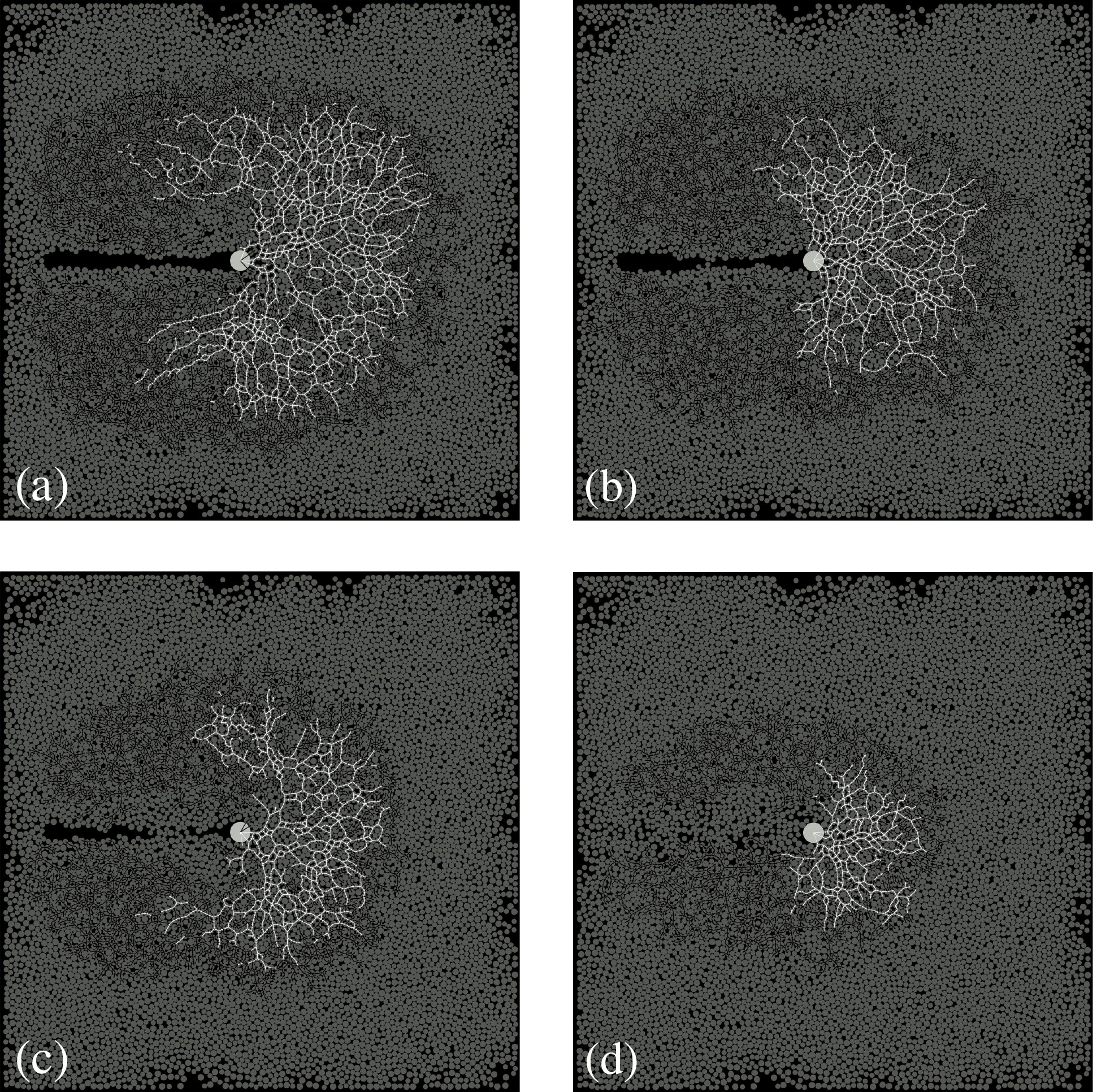}\\ 
	\end{center}
	\caption{Network of contact forces for different values of the coefficients of basal friction: (a) 100\%, (b) 25\%, (c) 10\% and (d) 0.1\% of the base values. Clear networks correspond to bearing (stronger) chains and darker networks to dissipative (weaker) chains, and all figures correspond to $t$  = 53.54 s.}
	\label{fig:network_basal_reduction}
\end{figure}

\begin{figure}[h!]
	\begin{center}
		\begin{minipage}{0.49\linewidth}
			\begin{tabular}{c}
				\includegraphics[width=0.90\linewidth]{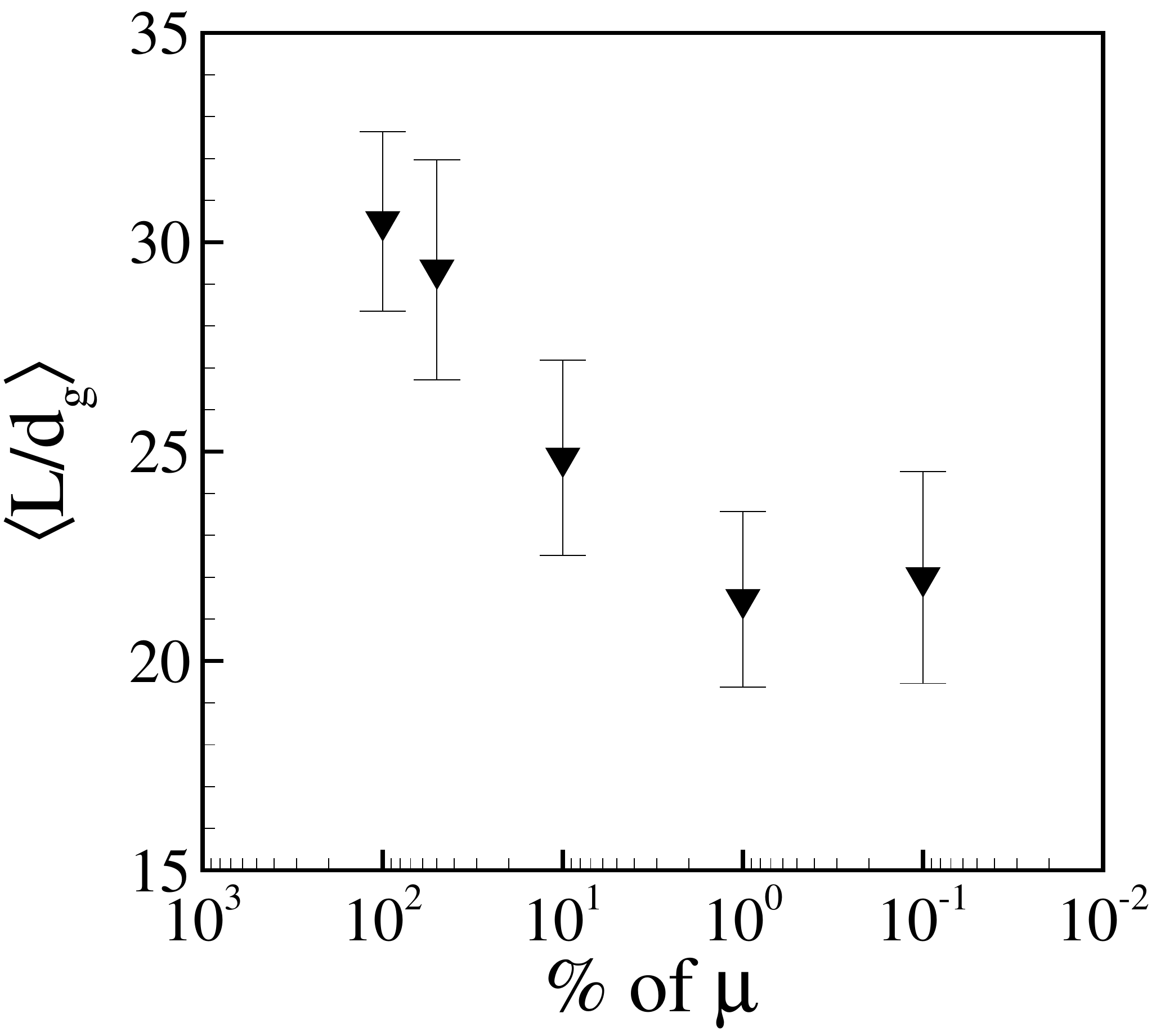}\\
				(a)
			\end{tabular}
		\end{minipage}
		\hfill
		\begin{minipage}{0.49\linewidth}
			\begin{tabular}{c}
				\includegraphics[width=0.90\linewidth]{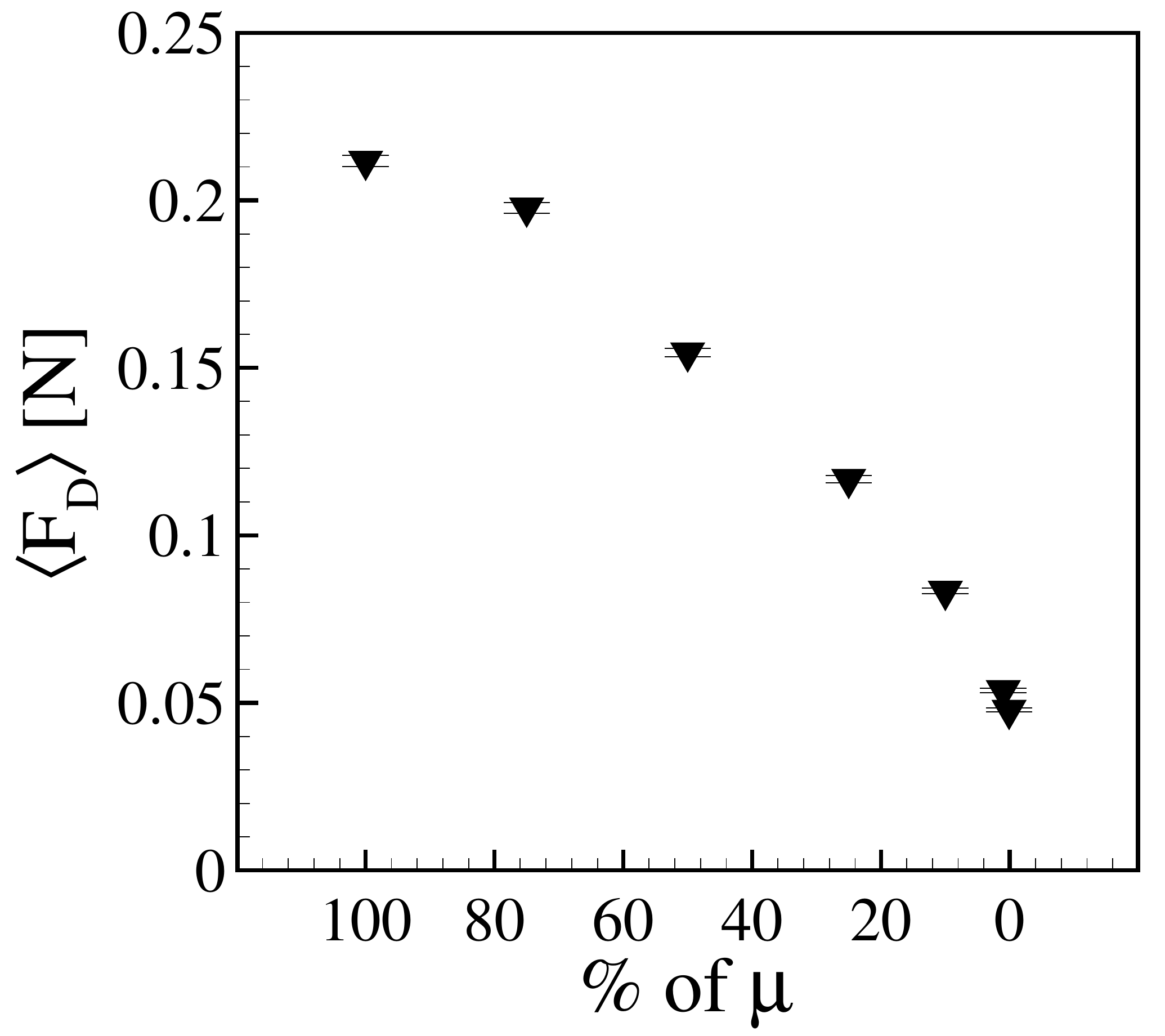}\\
				(b)
			\end{tabular}
		\end{minipage}
		\hfill
	\end{center}
	\caption{(a) Time-averaged length of chains $\left< L \right>$ normalized by $d_g$ and (b) time-averaged drag $\left< F_D \right>$ for different values of friction coefficients, in terms of percentages of the base values (see Tab. \ref{tabcoefficients_reduction}). In Fig. (a) error bars correspond to standard deviations and in Fig. (b) to standard errors.}
	\label{fig:mean_length_chains}
\end{figure}

Depending on the surface on which the disks move, the basal friction can be different. For instance, the friction can be much smaller over Teflon or ice than over acrylic, glass or metal alloys. The diversity of materials found both in nature and industry has thus motivated investigations of monolayers of particles sliding with different frictions. Kozlowski et al. \cite{Kozlowski} and Carlevaro et al. \cite{Carlevaro} investigated the effects of the basal friction of grains ($\mu_{gw}$ and $\mu_{s,gw}$) on the motion of an intruder with $\mu_{iw}$ = 0 in a Couette geometry. They found that two regimes of motion appear depending on the friction coefficients, and that in the case without basal friction chains occur only in front of the intruder during stick events. Although they advanced valuable information about the general behavior of the system, knowledge on how the chain dynamics varies with the basal friction is still missing.  

In the present Subsection, we inquire further into the effect of the basal friction on the network of contact forces. For that, we reduced the values of the coefficients of basal friction for both the intruder and grains by the same proportions, as indicated in Tab. \ref{tabcoefficients_reduction}: values of the static and dynamic coefficients, $\mu_{iw}$, $\mu_{gw}$ and $\mu_{s,gw}$, were reduced to 75, 50, 25, 10, 1, and 0.1\% of the original values (shown in Tab. \ref{tabcoefficients}). We analyze next how the density of contact networks, typical lengths of chains, and behavior of the cavity vary with the basal friction.

Figure \ref{fig:network_basal_reduction} shows the networks of contact forces for different values of the coefficients of basal friction, where Figs. \ref{fig:network_basal_reduction}a to \ref{fig:network_basal_reduction}d correspond to values of 100, 25, 10 and 0.1\% of the base value (Tab. \ref{tabcoefficients_reduction}). The figures show that the extents of both bearing and dissipative chains decrease with decreasing the basal friction. In order to quantify that, we computed the typical length $L$ of the bearing chains by measuring, along time, the maximum distance from the center of the intruder reached by bearing chains (see the Supplemental Material \cite{Supplemental} for a diagram showing how it is measured). Figure \ref{fig:mean_length_chains}a presents the time-averaged values of the typical length, $\left< L \right>$, normalized by the mean grain diameter $d_g$ for different values of friction coefficients (in terms of percentages of the base values). We observe that the extent of bearing chains decreases slightly with reducing the basal friction, the typical length decreasing by roughly 30\% when the basal friction is reduced from 0.4 and 0.7 to virtually 0 (0.0004 and 0.0007, respectively). As a result of the lower extent of load-bearing chains, the resultant drag on the intruder also decreases with decreasing the basal friction, Fig. \ref{fig:mean_length_chains}b showing that $\left< F_D \right>$ decreases one order of magnitude when basal frictions are reduced as before (the temporal evolution of $F_D$ for a basal friction of 0.1\% of the base value is available in the Supplemental Material \cite{Supplemental}, from which we observe a much lower level of fluctuations when compared with $F_D$ for the base value.)

\begin{figure}[h!]
	\begin{center}
		\includegraphics[width=0.50\columnwidth]{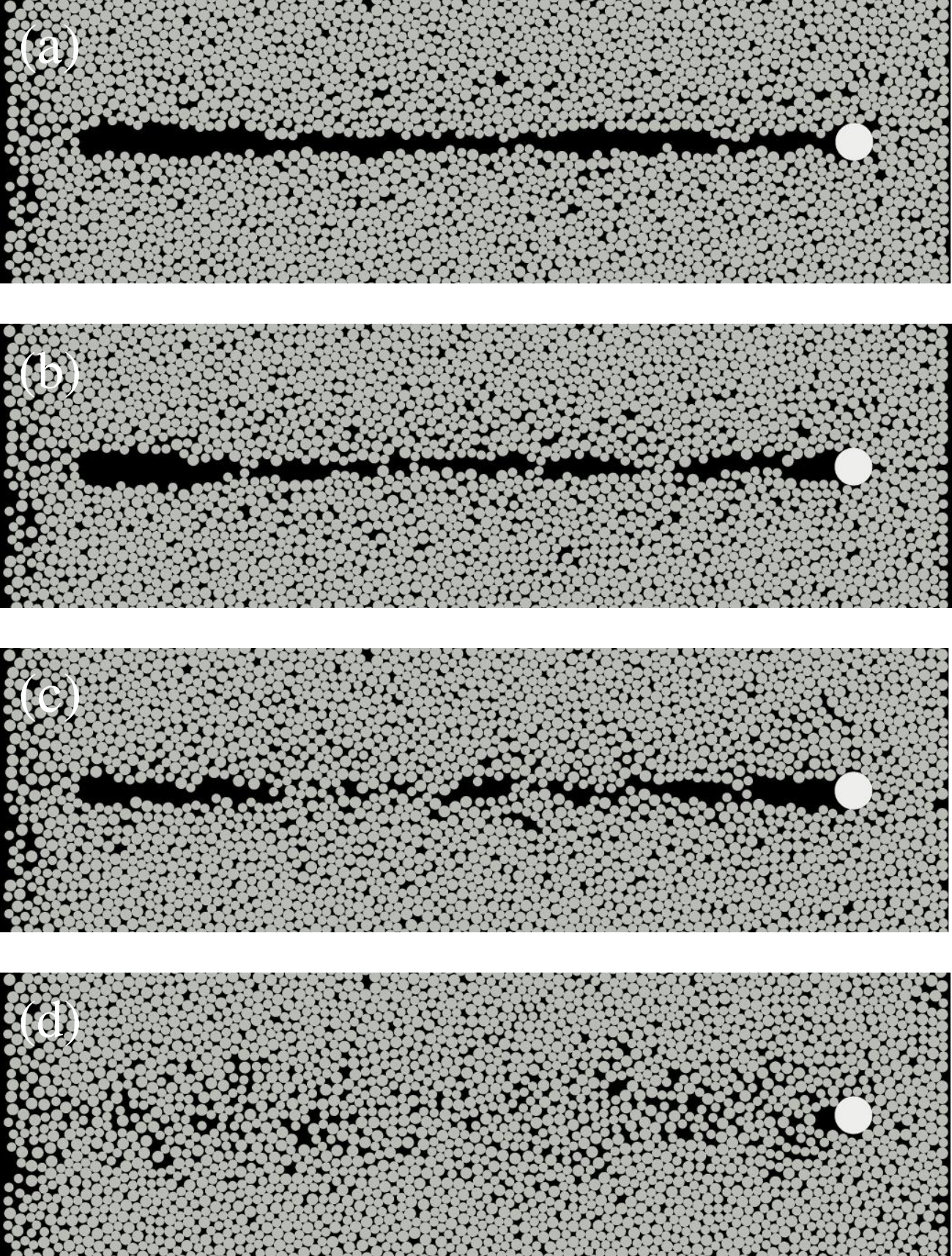}\\ 
	\end{center}
	\caption{Cavity for different basal frictions. Figures (a) to (d) correspond to 100, 25, 10 and 0.1\% of the base value (Tab. \ref{tabcoefficients_reduction}).}
	\label{fig:cavity_variation}
\end{figure}

The extent of bearing chains decreases under lower friction because in this case grains move easier when submitted to lower forces, breaking thus some of the chains. This is corroborated by the reductions of the downstream cavity as the basal friction decreases. Figure \ref{fig:cavity_variation} shows the cavity for different basal frictions, Figs. \ref{fig:cavity_variation}a to \ref{fig:cavity_variation}d corresponding to 100, 25, 10 and 0.1\% of the base value (Tab. \ref{tabcoefficients_reduction}). We observe a continuous reduction of the cavity size as the basal friction is reduced, with a very slight cavity (and wake) for the smallest value (Fig. \ref{fig:cavity_variation}d), whose size is comparable to that for higher packing fractions ($\phi$ $>$ 0.80) and 100\% of the basal friction (Fig. \ref{fig:setup}d).

\section{\label{sec:Conclu} CONCLUSIONS}

This paper investigated numerically the forces and structures in a two-dimensional granular system displaced by an intruder moving continuously. The granular system and the intruder consisted of 3D disks, all of them settled over a horizontal wall and confined by vertical walls, and, for the computations, we made use of the open-source DEM code LIGGGHTS \cite{Kloss, Berger} together with the DESIgn toolbox \cite{Herman}. By varying the intruder's velocity and the basal friction, we obtained the resultant force on the intruder and the instantaneous network of contact forces, which we analyzed at both the cell and grain scales. We first validated our numerical computations by replicating some of the experimental results obtained by Seguin et al. \cite{Seguin1}, and we afterward investigated the motion of particles and force transmission. We found that there is a bearing network that percolates large forces from the intruder toward the walls, being responsible for jammed regions and high values of the drag force, and a dissipative network that percolates small forces, in agreement with previous experiments on compressed granular systems. We then showed that anisotropy levels are higher for the bearing chains when compared with the dissipative ones, exhibiting some resemblance with shear jamming, and that anisotropy increases more in regions farther downstream of the intruder by the end of its motion, reaching values three times higher than those in upstream regions. We also found that the extent of the force network decreases with decreasing the basal friction, and that the void region (cavity) that appears downstream the intruder tends to disappear for lower values of basal friction. Finally, our results show that grains within the bearing chains creep while the chains break, revealing the mechanism by which bearing chains collapse, and allowing the intruder to proceed with its motion.

\section{\label{sec:Ack} ACKNOWLEDGMENTS}

The authors are grateful to FAPESP (Grant Nos. 2018/14981-7, 2019/20888-2 and 2020/04151-7) for the financial support provided.

\bibliography{references}

\end{document}